\documentclass[aps,pra,superscriptaddress,twocolumn,longbibliography]{revtex4-1}
\usepackage{graphicx, color, graphpap}
\usepackage{enumitem}
\usepackage{amssymb}
\usepackage{amsthm}
\usepackage{dsfont}
\usepackage{mathtools}
\usepackage[dvipsnames]{xcolor}
\usepackage{multirow}
\usepackage[colorlinks=true,citecolor=blue,linkcolor=magenta]{hyperref}
\usepackage[T1]{fontenc}
\usepackage{bbm}
\usepackage{thmtools,thm-restate}
\usepackage{verbatim}
\usepackage{titlesec}
\usepackage{amsmath}
\usepackage{graphicx}
\usepackage{float}
\usepackage{siunitx}
\usepackage{makecell}

\usepackage[ruled,vlined]{algorithm2e}

\newcommand{\bra}[1]{\langle#1|}

\newcommand{\ket}[1]{|#1\rangle}

\begin{document}

\title[]{Entanglement-variational Hardware-efficient Ansatz for Eigensolvers}

\author{Xin Wang}

\affiliation{
	Key Laboratory of Systems and Control, Academy of Mathematics and Systems Science, Chinese Academy of Sciences, Beijing 100190, P. R. China
}
\affiliation{
	University of Chinese Academy of Sciences, Beijing 100049, P. R. China
}

\author{Bo Qi}
\email{qibo@amss.ac.cn}
\affiliation{
	Key Laboratory of Systems and Control, Academy of Mathematics and Systems Science, Chinese Academy of Sciences, Beijing 100190, P. R. China
}
\affiliation{
	University of Chinese Academy of Sciences, Beijing 100049, P. R. China
}

\author{Yabo Wang}

\affiliation{
	Key Laboratory of Systems and Control, Academy of Mathematics and Systems Science, Chinese Academy of Sciences, Beijing 100190, P. R. China
}
\affiliation{
	University of Chinese Academy of Sciences, Beijing 100049, P. R. China
}
	
	\author{Daoyi Dong}

\affiliation{
	CIICADA Lab, School of Engineering, Australian National University, Canberra ACT 2601, Australia
}

\date{\today}

\begin{abstract}
  Variational quantum eigensolvers (VQEs) are one of  the most important and effective applications of quantum computing, especially in the current  noisy intermediate-scale quantum (NISQ) era. There are mainly two ways for VQEs: problem-agnostic and problem-specific. For problem-agnostic methods, they often suffer from trainability issues.  For problem-specific methods, their performance usually relies upon choices of initial reference states which are often hard to determine. In this paper, we propose an Entanglement-variational Hardware-efficient Ansatz (EHA), and  numerically  compare it with some widely used ansatzes by solving benchmark problems in quantum many-body systems and quantum chemistry. Our EHA is problem-agnostic and hardware-efficient, especially suitable for NISQ devices and having potential for wide applications. EHA can achieve a higher level of accuracy in finding  ground states and their energies in most cases even compared with problem-specific methods. The performance of EHA is robust to choices of initial states and  parameters initialization and it has the ability to quickly adjust the entanglement to the required amount, which is also the  fundamental reason for its superiority.
\end{abstract}

\maketitle

\section{Introduction}

Quantum computing has the potential to revolutionize many fields, including quantum many-body physics~\cite{huang2022provably,stetcu2022variational,morningstar2022simulation}, quantum chemistry~\cite{mcardle2020quantum,elfving2021simulating,cao2022progress}, material science~\cite{bauer2020quantum,ma2020quantum,bassman2021simulating}, and so on~\cite{egger2020quantum,cerezo2022challenges}. Among these, finding ground states and their corresponding energies of Hamiltonians is a fundamental problem~\cite{huang2022provably,stetcu2022variational,morningstar2022simulation,mcardle2020quantum,elfving2021simulating}. In the current  noisy intermediate-scale quantum (NISQ) era~\cite{Preskill2018quantumcomputingin,bharti2022noisyalgo,bharti2022noisyprog}, variational quantum eigensolvers (VQEs)~\cite{peruzzo2014variational,cerezo2021variational,tilly2022variational,kattemolle2022variational,li2023benchmarking} have been proposed  to solve this problem with the hope that approximate solutions could be found for large systems which are intractable with classical computers.

VQEs work in a hybrid quantum-classical manner~\cite{callison2022hybrid}. They employ a parameterized quantum circuit (PQC) to generate parameterized trial states and the variational parameters are updated by a classical optimizer through minimizing the objective function, which in general is the expectation value of the Hamiltonian with respect to the trial state.

To enhance  performance of VQEs, various methods have been developed from different perspectives. These include designing appropriate ansatzes for quantum circuits~\cite{ostaszewski2021structure,grimsley2019adaptive,tang2021qubit,ostaszewski2021reinforcement,du2022quantum,ding2022evolutionary,grimsley2023adaptive}, tailored initial states~\cite{google2020hartree,cade2020strategies}, proper parameter initializations~\cite{wang2023trainability,zhang2022escaping}, developing efficient optimization methods~\cite{Stokes2020quantumnatural}, employing feedback for iterations~\cite{magann2022feedback}, and utilizing classical post-processing with neural networks~\cite{zhang2022variational}, among others~\cite{stair2021simulating,cervera2021meta,fujii2022deep}. It is clear that the ansatz of a quantum circuit  directly determines the success of VQEs.   For instance, if the quantum circuit is poorly expressible  that cannot generate trial states close to the target state, then no other auxiliary methods can improve its performance~\cite{haug2021capacity,sim2019expressibility}. There are mainly two ways to design ansatzes for quantum circuits: problem-agnostic and problem-specific~\cite{cerezo2021variational}.

Hardware-efficient ansatz (HEA) is a well-known and widely used problem-agnostic method, which seeks to minimize the hardware noise by using native gates and connectives~\cite{kandala2017hardware,schuld2020circuit,choy2023molecular}.  When training quantum circuits based on HEAs, it often faces many challenges.
Shallow HEA circuits are poorly expressible, and may cause the landscape of the cost function to be swamped with spurious local minima under global measurements~\cite{bittel2021training,anschuetz2022quantum}. Deep circuits, however, will make the PQC too expressive resulting in barren plateaus (BPs)~\cite{mcclean2018barren,holmes2022connecting}, i.e., the cost gradient is exponentially small with the number of qubits and/or the circuit depth. Both of these issues will make the  PQC training extremely difficult.  A major reason is that in most of  existing HEAs, the entanglers are usually fixed, resulting in a lack of freedom to quickly adjust the entanglement of the trial states to the required amount. It is clear that the nature of the circuit ansatz determines the level of entanglement that can be achieved. On the one hand, generating a matched amount of entanglement is necessary to guarantee the convergency of eigensolvers based on HEAs. On the  other hand, while entanglement can usually be quickly generated within a few layers, the excess entanglement  cannot be removed efficiently~\cite{chen2022much}. It has been pointed out in~\cite{woitzik2020entanglement} that if the generated entanglement does not match the problem under study, it may hamper the convergence process. Moreover, it was argued in~\cite{marrero2021entanglement} that too much entanglement can result in BPs. In addition, it was stated in~\cite{leone2022practical} that even for shallow circuits, entanglement satisfying volume law should be avoided.

To address the training issues of PQCs, it has been suggested to design circuit ansatzes in a problem-specific manner. Examples include the Hamiltonian variational ansatz (HVA) also commonly referred to as a Trotterized  adiabatic state preparation ansatz~\cite{wiersema2020exploring}, and the hardware symmetry preserving ansatz proposed in~\cite{lyu2023symmetry}, which reduces the explored space of unitaries through symmetry preserving and is referred to as HSA in our paper.  In quantum computational chemistry, some chemically inspired ansatzes have been proposed by adapting classical  chemistry algorithms to run efficiently on quantum circuits~\cite{mcardle2020quantum}. The most notable one  is the unitary coupled cluster (UCC)~\cite{romero2018strategies} adapted from the coupled cluster (CC) method~\cite{bartlett2007coupled}. The variational UCC method is able to converge when used with multi-reference initial states. The UCC is usually truncated at the single and double excitations, known as UCCSD~\cite{romero2018strategies}. In similar spirit to UCCSD, another commonly used ansatz has been proposed in~\cite{arrazola2022universal}, which considers all single and double excitation gates acting on the reference state without flipping the spin of the excited particles, but where all gates are Givens rotations~\cite{arrazola2022universal}. We refer to this ansatz as Givens rotation with all singles and doubles (GRSD) in this paper. Moreover, adaptive ansatzes have been presented in~\cite{grimsley2019adaptive,tang2021qubit,farrell2023scalable}, where the structure of quantum circuits are optimized adaptively. It has been shown that  these adaptive ansatzes perform better  in terms of both circuit depth and chemical accuracy  than circuits that use parameters update alone.
For these problem-specific methods,  their performance usually  relies  upon choices of  initial reference states~\cite{cai2020resource,skogh2023accelerating,jattana2023improved}, which are often hard to determine.

In this paper, we present a hardware-efficient  ansatz for eigensolvers which allows for rapid adjustment of the entanglement to the required amount by making entanglers variational. Our ansatz is thus referred to as Entanglement-variational Hardware-efficient Ansatz (EHA). We validate its efficiency via numerical comparisons  with some widely used VQE ansatzes. By solving  benchmark problems in quantum many-body physics and quantum chemistry, our EHA has the following advantages: 1) It is hardware-efficient and can be applied to various kinds of problems, particularly suitable for NISQ devices. 2) In most of numerical experiments, EHA can attain a higher level of accuracy than other ansatzes, even as compared with problem-specific ansatzes. 3) For different choices of initial reference states and variational parameters, the performance of EHA is more robust as compared to other ansatzes. 4) The variational entangler design enables EHA to quickly adjust the entanglement to the desired amount, which is also the fundamental reason for its superiority.

This paper is organized as follows. In Section~\ref{sec:pre}, for subsequent comparison, we first introduce several basic models and corresponding ansatzes. Then we present our EHA, and demonstrate its advantages via numerical comparisons with other ansatzes in Section~\ref{sec:main}. Section~\ref{sec:conclusion} concludes the paper. Since we utilize many abbreviations,  for the convenience of reading, we summarize the frequently-used abbreviations and their full expressions in Table~\ref{tab:abbr}.

\begin{table}[h]
	\begin{ruledtabular}
	\begin{tabular}{cll}
		&Abbreviations  & Descriptions \\
		\hline		
		\hline
		\multicolumn{3}{c}{Quantum many-body model} \\
		\hline
		&HM & Heisenberg Model \\
		&TFIM &Transverse Field Ising Model \\
		&BHM & Bose Hubbard Model \\
		\hline
		\multicolumn{3}{c}{Variational Quantum Eigensolver(VQE) ansatz} \\
		\hline
		&EHA & Entanglement-variational \\
		&     &Hardware-efficient Ansatz\\
		&HEA & Hardware-Efficient Ansatz \\
		&HVA & Hamiltonian Variational Ansatz \\
		&HSA & Hardware Symmetry preserving Ansatz \\
		&GRSD & Givens Rotations with all \\
		&     & Single and Double excitations \\
		&UCCSD & Unitary Coupled Cluster with all\\
		&      &  Single and Double excitations \\
		&ADAPT-VQE & Adaptive Derivative-Assembled\\
		 &     &Pseudo-Trotter ansatz-VQE \\
	\end{tabular}
	\end{ruledtabular}
	\caption{\label{tab:abbr}Glossary.  }
\end{table}

\section{Preliminaries}
\label{sec:pre}

In this paper, we focus on the task of finding  ground eigenstates and their corresponding eigenenergies of Hamiltonians in quantum many-body physics and quantum chemistry. In this section, we introduce the models and ansatzes to  be used for comparison with our EHA.

\subsection{Quantum many-body models}

For quantum many-body physics, we first consider a 1-dimensional chain Heisenberg model (HM) consisting of $N$ spins~\cite{wiersema2020exploring,lyu2023symmetry}, whose Hamiltonian reads
\begin{equation}\label{XXZ}
\begin{aligned}
H_{\mathrm{HM}} = J\sum_{i=1}^{N-1} (\sigma^x_i \sigma^x_{i+1} + \sigma^y_i \sigma^y_{i+1} + \sigma^z_i \sigma^z_{i+1}),
\end{aligned}
\end{equation}
where $J=1$ sets the unit of energy,  $\sigma^x_i,~\sigma^y_i$ and $\sigma^z_i$ denote the Pauli $X$, $Y$ and $Z$ operators acting on the $i$-th qubit, respectively. The Hamiltonian Eq.~(\ref{XXZ}) describes a model of interacting systems that cannot be mapped to free fermions. It supports symmetries including the conservation of the spin components in all directions, i.e., $[H_{\mathrm{HM}}, S_{\alpha}]=0$ with $S_{\alpha} = \frac{1}{2} \sum_{i}^{ } \sigma_i^{\alpha}$ for $\alpha = x,~y$ and $z$, as well as the total spin, namely $[H_{\mathrm{HM}}, S_{\mathrm{tot}}^2]=0$ with $S_{\mathrm{tot}}^2 = S_x^2 + S_y^2 + S_z^2$.

We also consider free fermionic systems described by the transverse field Ising model (TFIM)~\cite{wiersema2020exploring,lyu2023symmetry}, whose Hamiltonian reads
\begin{equation}\label{TFIM}
\begin{aligned}
H_{\mathrm{TFIM}} = J_z \sum_{i=1}^{N-1} \sigma^z_i \sigma^z_{i+1} + h_x \sum_{i=1}^{N} \sigma^x_i,
\end{aligned}
\end{equation}
where $J_z$ represents the exchange coupling and $h_x$ depicts the strength of the transverse magnetic field. For this Hamiltonian, it is well-known that a quantum phase transition occurs at $J_z=h_x$, and at this critical point, the ground state is highly entangled and in a complex form~\cite{lyu2023symmetry}.

\subsection{Hardware-efficient ansatzes}
In the NISQ era, VQEs have been proposed for eigensolvers with the hope to demonstrate potential quantum advantages when dealing with large systems which are beyond the power of classical computers.

Given a Hamiltonian $H$, when employing VQEs, we first need to design an ansatz to build a PQC
$U(\boldsymbol{{\theta}})$ with variational parameters $\boldsymbol{\theta}$. Staring from a given initial reference state $\ket{\psi_0}$, we use the generated trial state $\ket{\psi(\boldsymbol{{\theta}})}=U(\boldsymbol{\theta})\ket{\psi_0}$ to approximate the ground state of
$H$. The variational parameters $\boldsymbol{\theta}$  are adaptively updated   by a classical optimizer via minimizing the expectation of  $H$ at the trial state, which reads
\begin{equation}
	\label{eq:Cost}
C(\boldsymbol{\theta})  = \operatorname{Tr} \left[ H U(\boldsymbol{\theta}) \ket{\psi_0} \bra{\psi_0} U^{\dagger}(\boldsymbol{\theta})  \right].
\end{equation}

\begin{figure}[htpb]
	\centering
	\includegraphics[width=0.45\textwidth]{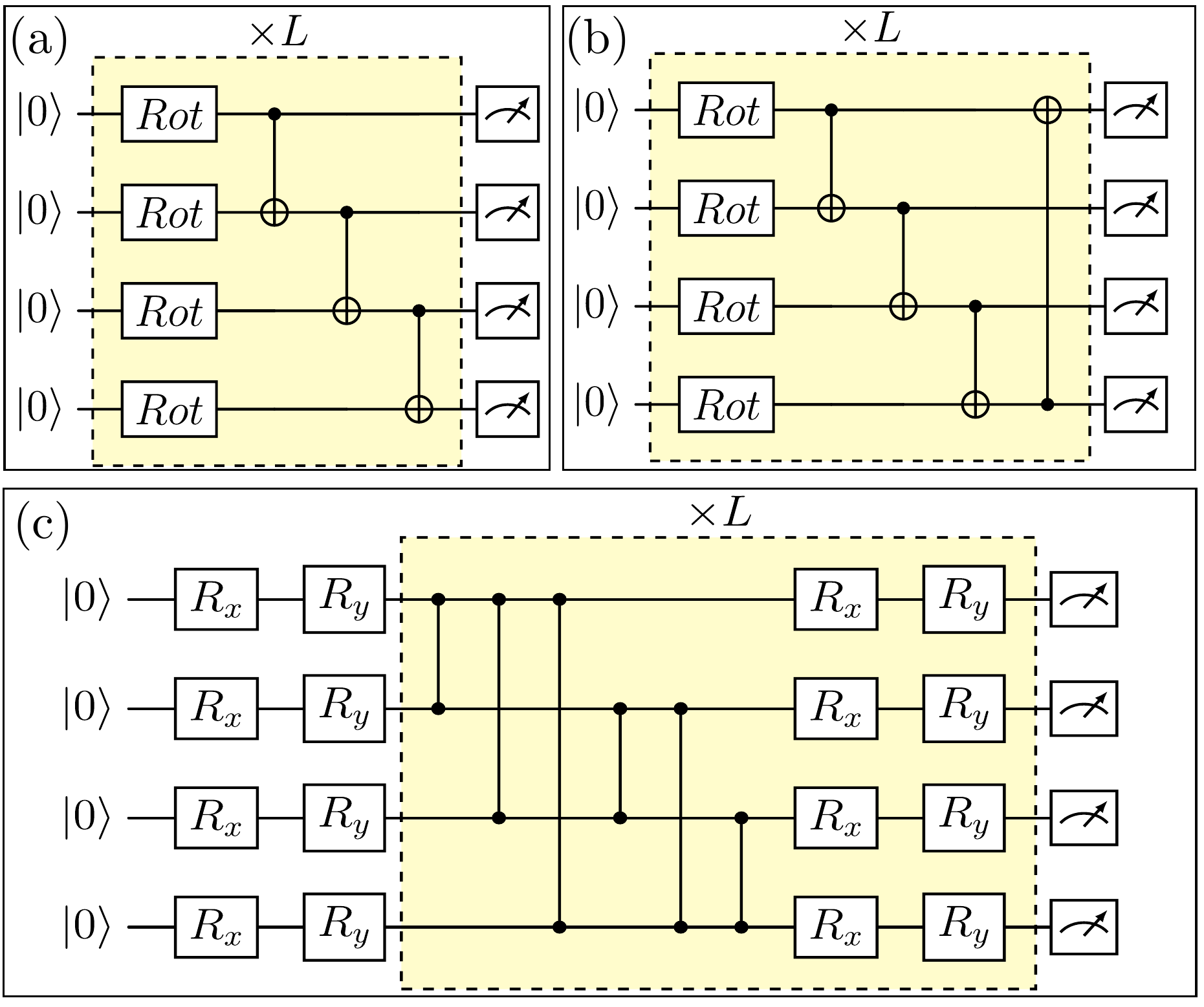}
	\caption{Quantum circuits for HEAs. (a) CX-line,  whose single-qubit modules are $Rot$ gates defined in Eq.~(\ref{rot}), and the entangling gates are CX gates arranged in a line pattern.~(b) CX-ring, whose entangling gates are arranged in a ring pattern.~(c) CZ-complete, whose single-qubit modules are $R_yR_x$ gates, and the entangling gates are CZ gates arranged in a complete pattern. The yellow shaded circuits are repeated for $L$ times. }
	\label{fig:HEA_ansatz}
\end{figure}

To minimize the hardware noise in the NISQ era, HEAs have been widely used for eigensolvers.  Three commonly used HEAs are illustrated in Fig.~\ref{fig:HEA_ansatz}, and will be compared with our EHA. The initial reference state of these three HEAs is usually set to be $\rho_0 = \ket{0}\bra{0}$. For the circuit in Fig.~\ref{fig:HEA_ansatz}(a), the single-qubit rotations take the form of
\begin{equation}\label{rot}
Rot(\phi, \theta, \omega)=R_z(\omega) R_y(\theta) R_z(\phi),
\end{equation}
with $R_{z}(\phi)=e^{-i \phi Z / 2}$, $R_{y}(\theta)=e^{-i \theta Y / 2}$, and the entangling gates are CX gates arranged in a line pattern~\cite{choy2023molecular}. Thus, we refer to it as CX-line. The ansatz in Fig.~\ref{fig:HEA_ansatz}(b) is named after CX-ring, since its single-qubit modules and entangling gates are the same as those in CX-line, except that the entangling gates are in a ring pattern~\cite{schuld2020circuit}. For the ansatz in Fig.~\ref{fig:HEA_ansatz}(c), its single-qubit rotations are in the form of  $R_y(\theta)R_x(\phi)$ with $R_{x}(\phi)=e^{-i \phi X / 2}$~\cite{zhang2022escaping}. Since its entangling gates are CZ gates arranged in a complete pattern~\cite{kim2021universal}, we call it CZ-complete.

\subsection{Problem-specific ansatzes}
The problem-agnostic HEAs often suffer from difficult training issues. In addition, HEAs do not preserve any symmetry in general. In this subsection, we briefly introduce two important problem-specific ansatzes: Hamiltonian variational ansatz (HVA)~\cite{wiersema2020exploring} and hardware symmetry preserving ansatz (HSA)~\cite{lyu2023symmetry}.

{\bf HVA:}~Assume   the Hamiltonian $H$ can be decomposed into a summation with a  total number of $S$ terms as
\begin{equation}\label{H}
H = \sum_{s=1}^{S} H_s.
\end{equation}
Then we can construct  the HVA  in the form of
\begin{equation}\label{U-hva}
\begin{aligned}
U(\boldsymbol{\theta}) = \prod_{l=1}^{L} \left[ \prod_{s=1}^{S} \exp (-i \theta_{l,s} H_s)  \right].
\end{aligned}
\end{equation}
Specifically, for an $N$-qubit HM described by Eq.~(\ref{XXZ}), when the qubit number is even, its Hamiltonian can be decomposed into~\cite{wiersema2020exploring}
\begin{equation*}
	H_{\mathrm{HM}}=H^{\text {even }}+H^{\text {odd }},
\end{equation*}
with
\begin{equation*}
\begin{aligned}
H^{\text {even }} & =H_{x x}^{\text {even }}+H_{y y}^{\text {even }}+H_{z z}^{\text {even }},\\
H^{\text {odd }} & =H_{x x}^{\text {odd }}+H_{y y}^{\text {odd }}+H_{z z}^{\text {odd }},
\end{aligned}
\end{equation*}
where  for $\alpha = x,~y$ and $z$, $H_{\alpha \alpha}^{\mathrm{even}}=\sum_{i=1}^{N/2} \sigma_{2 i-1}^{\alpha} \sigma_{2 i}^{\alpha}$ and $H_{\alpha \alpha}^{\mathrm{odd}}=\sum_{i=1}^{N/2  - 1} \sigma_{2 i}^{\alpha} \sigma_{2 i+1}^{\alpha}.$ Then   $U_{\mathrm{HM}}(\boldsymbol{\theta},\boldsymbol{\phi},\boldsymbol{\beta},\boldsymbol{\gamma})$ reads
\begin{equation*}
\begin{aligned}
&U_{\mathrm{HM}}(\boldsymbol{\theta},\boldsymbol{\phi},\boldsymbol{\beta}, \boldsymbol{\gamma})=  \prod_{l=1}^{L} \Big{[} G\left(\gamma_{l}, H_{x x}^{\text {even }}\right) G\left(\gamma_{l}, H_{y y}^{\text {even }}\right) \cdot \\
&G\left(\beta_{l}, H_{z z}^{\text {even }}\right)  G\left(\phi_{l}, H_{x x}^{\text {odd }}\right)  G\left(\phi_{l}, H_{y y}^{\text {odd }}\right) G\left(\theta_{l}, H_{z z}^{\text {odd }}\right) \Big{]},
\end{aligned}
\end{equation*}
with $G(x, A)=e^{-i \frac{x}{2} A}$.

\begin{figure}[htpb]\label{XXYYZZ}
	\centering
	\includegraphics[width=0.48\textwidth]{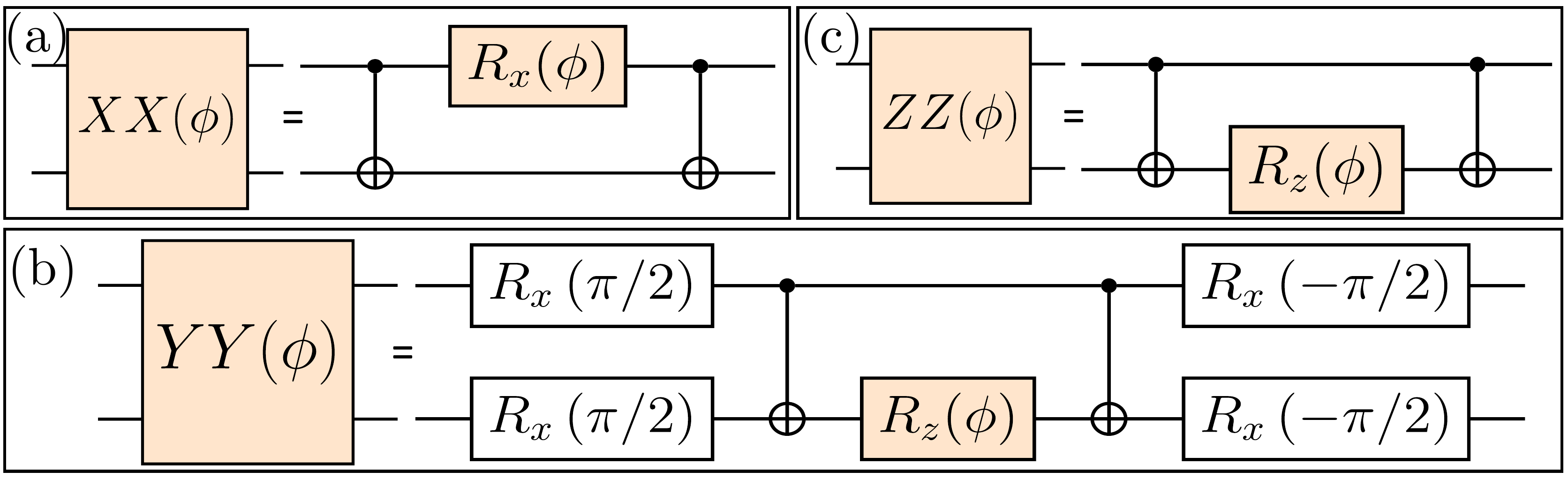}
	\caption{Hardware-efficient circuits for entangling gates. (a) Circuit for $XX$.~(b)~Circuit for $YY$.~(c)~Circuit for $ZZ$.}
	\label{fig:gate}
\end{figure}

The whole circuit for $U_{\mathrm{HM}}(\boldsymbol{\theta},\boldsymbol{\phi},\boldsymbol{\beta},\boldsymbol{\gamma})$ can be found in Fig.~\ref{fig:HVA_XXZ} in Appendix~\ref{AA}. To implement $U_{\mathrm{HM}}(\boldsymbol{\theta},\boldsymbol{\phi},\boldsymbol{\beta},\boldsymbol{\gamma})$, the entangling operators,
$XX(\phi)=e^{-i \frac{\phi}{2}(X \otimes X)}$, $YY(\phi)=e^{-i \frac{\phi}{2}(Y \otimes Y)}$ and $Z Z(\phi)=e^{-i \frac{\phi}{2}(Z \otimes Z)}$, are employed whose circuits are illustrated in Fig.~\ref{fig:gate}.

Note that the initial reference state plays a crucial role in problem-inspired ansatzes~\cite{google2020hartree,cade2020strategies}, which is quite different from the case in HEAs. To guarantee the efficiency of HVA for HM, the initial reference state is chosen to be ${\otimes ^{\frac{N}{2}}} {\ket{\Psi^{-}}}$ with $|\Psi^{-}\rangle=\frac{1}{\sqrt{2}}(|01\rangle-|10\rangle)$.

As for an $N$-qubit TFIM with $J_z=h_x=-1$ and $N$ being even,  the  Hamiltonian  $H_{\mathrm{TFIM}}$ can be decomposed into
\begin{equation}\label{TFIMequal}
\begin{aligned}
H_{\mathrm{TFIM}}=-\sum_{i=1}^{N-1} \sigma^z_i \sigma^z_{i+1} -\sum_{i=1}^{N} \sigma^x_i  = H_{zz} + H_x,
\end{aligned}
\end{equation}
with $H_{zz} =  -\sum_{i} \sigma^z_i \sigma^z_{i+1}$ and $H_x = -\sum_{i}\sigma^x_i$. Then the  HVA  $U_{\mathrm{TFIM}}(\boldsymbol{\beta},\boldsymbol{\gamma})$ reads

\begin{equation*}
\begin{aligned}
U_{\mathrm{TFIM}}(\boldsymbol{\beta},\boldsymbol{\gamma}) = \prod_{l=1}^{L} \left[ \exp(- i \frac{\gamma_{l}}{2} H_x ) \exp( -i \frac{\beta_l}{2} H_{zz})  \right],
\end{aligned}
\end{equation*}
whose circuit diagram is illustrated in Fig.~\ref{fig:TFIM_HVA_HSA}(a) in Appendix~\ref{AA}. For TFIM, the initial reference state is chosen to be $ {\otimes^{N}}{\ket{+}}$ with $\ket{+} =\frac{1}{\sqrt{2}}(|0\rangle+|1\rangle)$.

{\bf HSA:}~In~\cite{lyu2023symmetry},  a HSA was proposed to improve the performance of eigensolvers by exploiting the symmetries of the Hamiltonian.

 To be specific, if the ansatz $U(\boldsymbol{\theta})$ satisfies $[U(\boldsymbol{\theta}),S_z] = 0$, then it conserves the number of  excitation. While if $[U(\boldsymbol{\theta}),S_{\mathrm{tot}}^2] = 0$, then it preserves the total spin. To realize HSA, the following  more complex entangling gate~\cite{vatan2004optimal}
 \begin{equation*}
\begin{aligned}
	\mathcal{N}(\theta,\phi,\beta) =e^{i\left(\theta\sigma^x_1 \sigma^x_2+\phi\sigma^y_1 \sigma^y_2+\beta\sigma^z_1 \sigma^z_2\right)}.
\end{aligned}
\end{equation*}
has been employed. The circuit of realizing  $\mathcal{N}(\theta,\phi,\beta)$ is shown in Fig.~1(b) in~\cite{lyu2023symmetry} (Fig.~\ref{fig:HSA_XXZ}(b) in Appendix~\ref{AA}).

As for the HM in  Eq.~(\ref{XXZ}), a single block circuit for realizing $S_z$-conserving and $S_{\mathrm{tot}}$-conserving ansatz is shown in Fig.~1(c) and Fig.~1(d) in~\cite{lyu2023symmetry}, respectively. For ease of reading, we illustrate the HSA circuit in Fig.~\ref{fig:HSA_XXZ} in Appendix~\ref{AA}.  Since the ground state of the Hamiltonian Eq.~(\ref{XXZ}) is a global singlet with both the total spin $s = 0$ and the spin component $s_z =0$, the initial reference state is chosen to be ${\otimes^ {\frac{N}{2}}} {\ket{\Psi^{-}}}$ to guarantee the preservation of these symmetries in trial states.

The authors in~\cite{lyu2023symmetry} also generalized HSA to solve TFIM, whose circuit is shown in Fig.~7(a) in~\cite{lyu2023symmetry} (Fig.~\ref{fig:TFIM_HVA_HSA}(b) in Appendix~\ref{AA}). The initial reference state is chosen to be ${\otimes^N}|+\rangle$.

\subsection{Quantum chemistry models and ansatzes}

Solving the low lying energies of the electrons in molecules has attracted significant attention~\cite{mcardle2020quantum} since it was first introduced by~\cite{aspuru2005simulated} in the context of quantum computational chemistry. It is often a starting point for  more complex analysis in chemistry, including the calculation of reaction rates, the determination of molecular geometries and thermodynamic phases, among others~\cite{helgaker2012recent}.

In this paper, we try to find the ground eigenstates and eigenenergies of the electronic Hamiltonian of $\mathrm{H}_3^{+}$ cation, $\mathrm{HF}$, $\mathrm{H}_5$, $\mathrm{BeH_2}$  and $\mathrm{LiH}$ molecules. For conciseness, we work in atomic units, where the length unit is 1 $\si{\angstrom}$ (1 $\si{\angstrom}=1\times 10^{-10}$ m), and the energy unit is 1 Hartree (1 Hartree = $27.211$ eV).  We utilize the second quantized representation to simulate chemical systems on a quantum computer~\cite{babbush2016exponentially}. To do this, we need to select a basis set,  which is used to approximate the spin-orbitals of the investigated molecule.  A suitably large basis set is crucial for obtaining accurate results. In this paper, we take the Slater type orbital-$3$ Gaussians (STO-$3$G)~\cite{helgaker2013molecular} as the spin-orbital basis for second quantization. Then we utilize the occupation number basis to represent whether a spin-orbital is occupied. Next we can employ the Jordan-Wigner encoding~\cite{nielsen2005fermionic} to map the second quantized fermionic Hamiltonian into a linear combination of Pauli strings, each of which is a product of single qubit Pauli operators. More details can be found in~\cite{mcardle2020quantum}.

While it is important to understand the whole procedure of how to map electronic structure problems onto a quantum computer, every step from selecting a basis to producing an encoded qubit Hamiltonian can be carried out using a quantum computational chemistry package such as \texttt{OpenFermion}~\cite{mcclean2020openfermion} and \texttt{PennyLane}~\cite{bergholm2018pennylane}. In this paper, we adopt \texttt{PennyLane} to  implement  UCCSD and GRSD.

\section{Main Results}
\label{sec:main}

In this section, we first present our hardware-efficient ansatz having variational entangling gates for eigensolvers. We then compare it with the  ansatzes introduced by focusing on  models introduced in Section~\ref{sec:pre}.

\subsection{Entanglement-variational hardware-efficient ansatz }

\begin{figure}
	\centering
	\includegraphics[width=0.45\textwidth]{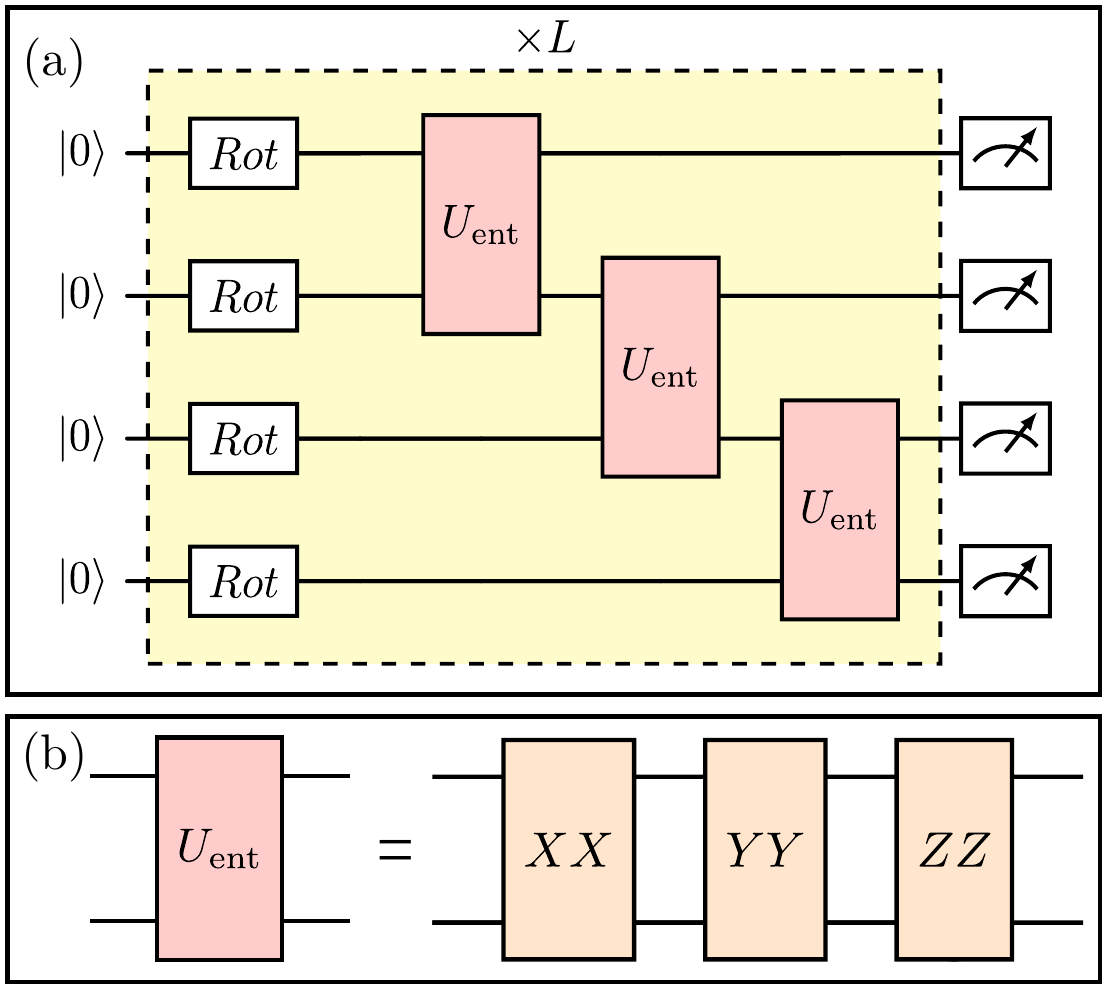}
	\caption{Quantum circuit for EHA. (a) The circuit consists of $L$ blocks, and each block is composed of a layer of single qubit rotational gates defined by Eq.~(\ref{rot}) and variational entanglers $U_{\mathrm{ent}}$ arranged in a line pattern. (b) Each variational entangler $U_{\mathrm{ent}}$ is composed of $XX$, $YY$ and $ZZ$ gates in series, whose realizations are illustrated in Fig.~\ref{fig:gate}. The variational parameters of the entanglers are (in general) different. The initial reference state is set to be $\otimes^n|0\rangle$.}
	\label{fig:QERA}
\end{figure}

Our aim is to design a hardware-efficient ansatz for  eigensolvers  which can be efficiently applied to solve various kinds of Hamiltonians for different systems.

To approximate  target ground states, it is imperative to generate trial states having a matched amount of entanglement.  However,  there is a lack of freedom to adjust  entanglement to the required level in most of existing HEAs, as their entanglers are fixed. The entanglement can usually be generated within a few layers, however, once the entanglement is in excess of requirement, it cannot be removed efficiently. The excess entanglement will result in too much expressibility leading to the phenomenon of BPs, which makes the training extremely difficult and greatly hampers the efficiency of HEAs.

Note that the entanglement is generated through  entangling gates. Therefore, to improve the ability of regulating the generated entanglement of HEAs, a natural idea is to make  entangling gates tunable with variational parameters. Based on this, we present our entanglement-variational hardware-efficient ansatz (EHA) as illustrated in~Fig.~\ref{fig:QERA}.

It is clear that the entanglers $$U_{\mathrm{ent}}(\boldsymbol{\theta_{l,i}})=ZZ(\theta_{l,i,3})YY(\theta_{l,i,2})XX(\theta_{l,i,1})$$ are tunable, which is the most significant difference from existing HEAs. Our ansatz is hardware-efficient, since the variational $XX$, $YY$ and $ZZ$ gates  can be realized by native single qubit gates and CX gates as illustrated in Fig.~\ref{fig:gate}. Moreover, we would like to point out that since the entanglers $XX$, $YY$ and $ZZ$ are native to hardwares themselves in some quantum computers~\cite{debnath2016demonstration}, this can further significantly improve the efficiency of our EHA.

Although HVA and HSA also employ variational $XX$, $YY$ and $ZZ$ gates to implement entangling operations, the principle of their design is rather different from our EHA. There they first decompose the Hamiltonian of a specific problem into components as in Eq.~(\ref{H}), and then design variational circuits to realize each component Hamiltonian as described by Eq.~(\ref{U-hva}). Note that in quantum many-body physics, Hamiltonians are described in terms of Pauli strings. Thus, when realizing the component Hamiltonians, variational gates, e.g., $XX$, naturally appear. Their constructed circuits are problem-specific. However, our EHA is problem-agnostic and can be applied to various problems. Moreover, it is worth pointing out that for the HM,  although the circuits of  EHA  (Fig.~\ref{fig:QERA}), HVA (Fig.~\ref{fig:HVA_XXZ}) and HSA (Fig.~\ref{fig:HSA_XXZ}) appear similar apart from the way the entangling gates are arranged, there is an additional layer of single-qubit rotational gates with each having three variational parameters as described by Eq.~(\ref{rot}) for each block.  For the TFIM, compared to HVA and HSA (Fig.~\ref{fig:TFIM_HVA_HSA}), each block of EHA (Fig.~\ref{fig:QERA}) has more degrees of freedom in adjusting the entangling and single-qubit variational gates.
It has been demonstrated in~\cite{larocca2023theory} that by increasing the number of variational parameters, local minima of the cost function can be transformed into saddle points. Furthermore,  it has been shown that gradient descent only converges to minima rather than saddle points~\cite{lee2016gradient}. Thus, the additional degrees of variational parameters in our EHA can help enhance the performance of eigensolvers compared to HVA and HSA. This will be validated in the following subsections.

We now compare our EHA with the ansatzes introduced in Section~\ref{sec:pre} in finding the ground states and their energies of Hamiltonians of various kinds of systems. In this paper, the actual ground states and their energies of Hamiltonians are  obtained by employing the classical method \texttt{NumPy}~\cite{harris2020array}. We demonstrate that  as compared with other ansatzes, our EHA can approximate the target ground states with a higher level of accuracy in most cases, and its performance is more robust with respect to different initializations and different initial reference states. In addition, our EHA can quickly adjust the entanglement of trial states to the required amount.

\subsection{Higher level of accuracy and robustness}\label{sec:eff}

\subsubsection{Quantum many-body problems}

In the first part of Subsection~\ref{sec:eff}, we focus on finding eigenstates of a $12$-qubit HM in Eq.~(\ref{XXZ}) and an 12-qubit TFIM in Eq.~(\ref{TFIM}).

Since 2-qubit entangling gates  are  valuable resources in quantum computing, when comparing the performance of different ansatzes, to be fair,  we ensure that they employ roughly the same number of basic 2-qubit entangling gates. As an illustration, when solving an $N$-qubit HM ($N$ is even), as shown in Fig.~\ref{fig:gate} and Fig.~\ref{fig:QERA}, our EHA utilizes a total number of $6(N-1)$ CX gates per block. While for each block, there are $(N-1)$ CX gates for  CX-line, $N$ CX  gates for CX-ring, $\frac{N(N-1)}{2}$ CZ gates for CZ-complete, $6(N-1)$ CX gates for  HVA, and $3(N-1)$ CX gates for  HSA. Here, we do not include the CX gates utilized to prepare the initial reference states for  HVA and HSA. Thus, if there are $L$ blocks in our EHA, then the number of blocks should be $6L$ for  CX-line, $\left\lceil\frac{6L(N-1)}{N} \right\rceil$  for CX-ring,  $\left\lceil \frac{12L}{N} \right\rceil $ for CZ-complete,  $L$ for HVA and $2L$ for HSA. Here, $\left\lceil \cdot \right\rceil$ denotes the roundup function.

As for the initialization of variational parameters, for CZ-complete, we use Gaussian initialization whose mean is 0 and variance is $1/L$ as mentioned in~\cite{zhang2022escaping}. While for the other ansatzes, we adopt  the uniform distribution $\mathcal{U}[-\pi,\pi]$. We employ the Adam optimizer provided by  \texttt{PennyLane} for training, and the step size is set to be $0.01$ for all ansatzes, unless otherwise stated.  Each ansatz is implemented  $10$ times, and for each realization the initial parameters are drawn according to  the assumed distribution. In the following figures, otherwise stated, the black dashed line denotes the actual ground state energy of the considered case, which serves as a baseline for comparison. All the other lines indicate the respective average value of the expectation of the considered Hamiltonian over $10$ realizations for different ansatzes. The shaded areas represent the smallest area that includes all the  behaviors of the 10 realizations.
 Here, for simplicity, we do not take into account of the impact of limited measurement shots, which is a limiting factor for NISQ devices. We also perform additional experiments under limited shot counts, and the results demonstrate that our EHA has similar advantages to the ideal case (infinite shots) compared to other ansatzes. We illustrate one of the results in Appendix \ref{app:shots}.

\begin{figure}[htpb]
	\centering
	\includegraphics[width=0.45\textwidth]{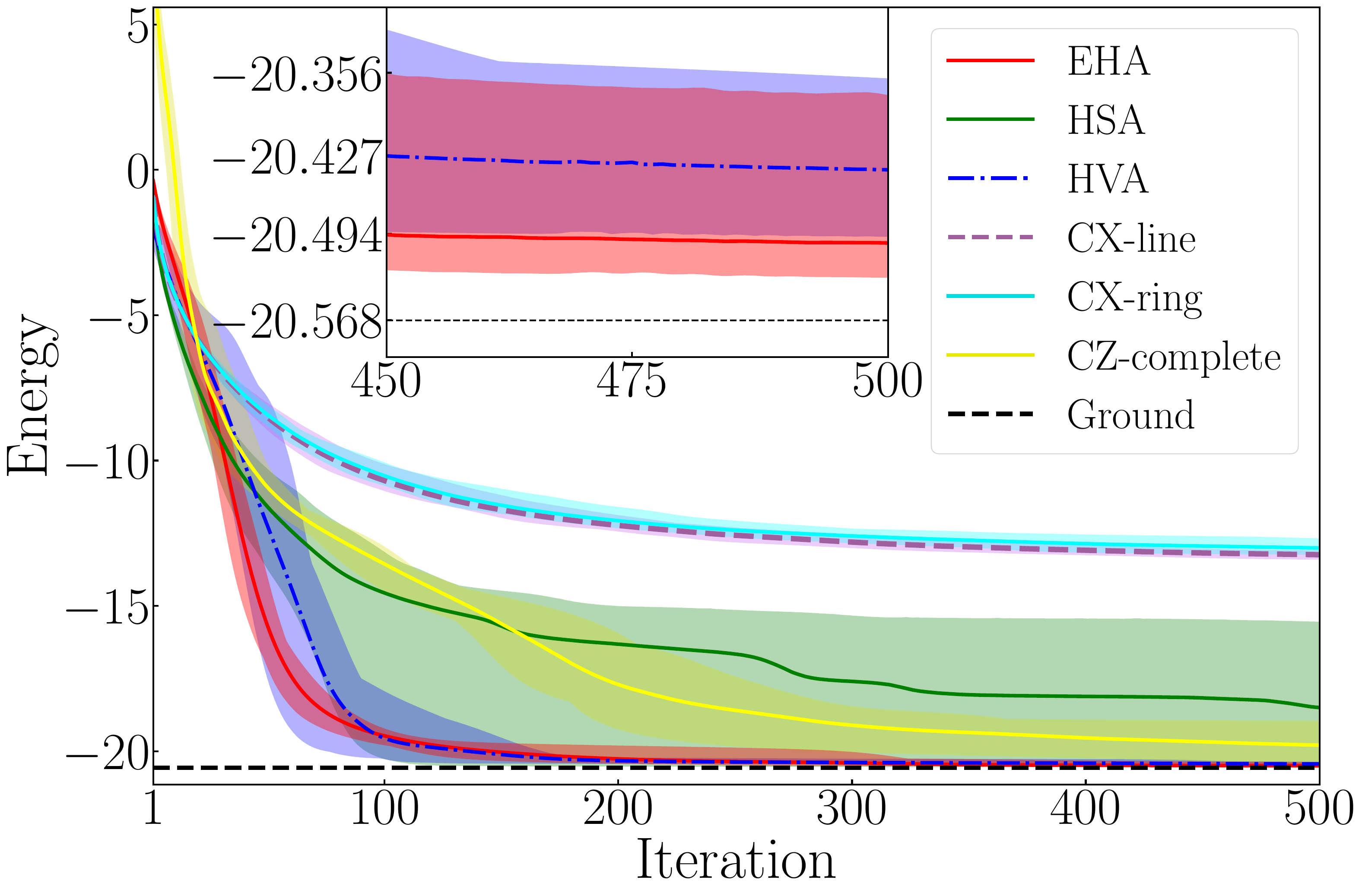}
	\caption{Comparison of our EHA with other ansatzes for HM in Eq.~(\ref{eq:XXZ}). The inset illustrates the fine difference between EHA and HVA. Our EHA is superior to other ansatzes in obtaining a lower energy, and its performance is robust with respect to different realizations. }
	\label{fig:res_XXZ}
\end{figure}

To compare, we first consider a $12$-qubit HM, whose Hamiltonian reads
\begin{equation}
	\label{eq:XXZ}
\begin{aligned}
H_{\mathrm{HM}} = \sum_{i=1}^{N} (\sigma^x_i \sigma^x_{i+1} + \sigma^y_i \sigma^y_{i+1} + \sigma^z_i \sigma^z_{i+1}),
\end{aligned}
\end{equation}
with $N = 12$. The numbers of blocks in our EHA, CX-line, CX-ring, CZ-complete, HVA and HSA are 10, 60, 55, 10, 10 and 20, respectively. We illustrate their performance in Fig.~\ref{fig:res_XXZ}.   From the perspective of the average energy over different realizations, it is clear that our EHA outperforms all the other ansatzes. Among them, the performances of CX-ring and CX-line are very poor. The performance of HSA has an obvious gap with our EHA. In addition, our EHA is robust to different parameter initializations, since all the 10 realizations approximately converge to the actual ground state energy. In contrast, the performance of HSA is very sensitive to parameter initializations.

\begin{figure}[htpb]
	\centering
	\includegraphics[width=0.45\textwidth]{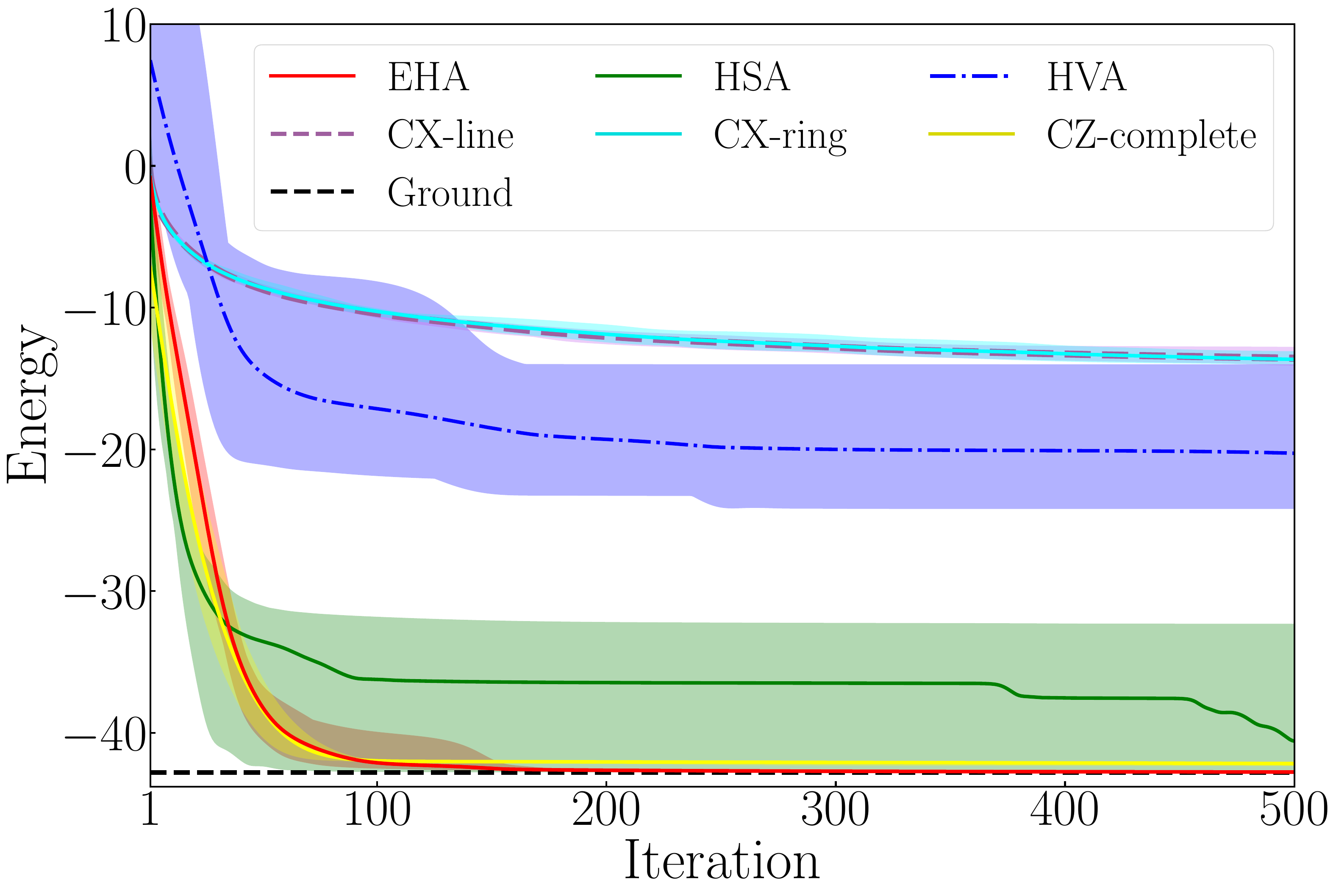}
	\caption{Comparison of our EHA with other ansatzes for  the  TFIM in Eq.~(\ref{eq:TFIM}).  Our EHA is much better than the other ansatzes, as it can robustly obtain a lower energy.}
\label{fig:TFIM_cost}
\end{figure}

Next, for the  $12$-qubit TFIM, we first consider a Hamiltonian which reads
\begin{equation}
	\label{eq:TFIM}
\begin{aligned}
H_{\mathrm{TFIM1}} = - \sum_{i=1}^{N-1} \sigma^z_i \sigma^z_{i+1} + 3.5 \sum_{i=1}^{N}   \sigma^x_i,
\end{aligned}
\end{equation}
with $N = 12$. The numbers of blocks in our EHA, CX-line, CX-ring, CZ-complete, HVA, and HSA are 4, 24, 22, 4, 4 and 8, respectively. The results are illustrated in Fig.~\ref{fig:TFIM_cost}.  We find that our EHA  outperforms all  the other ansatzes, as it can robustly attain a lower energy.

\begin{figure}[htpb]
	\centering
	\includegraphics[width=0.42\textwidth]{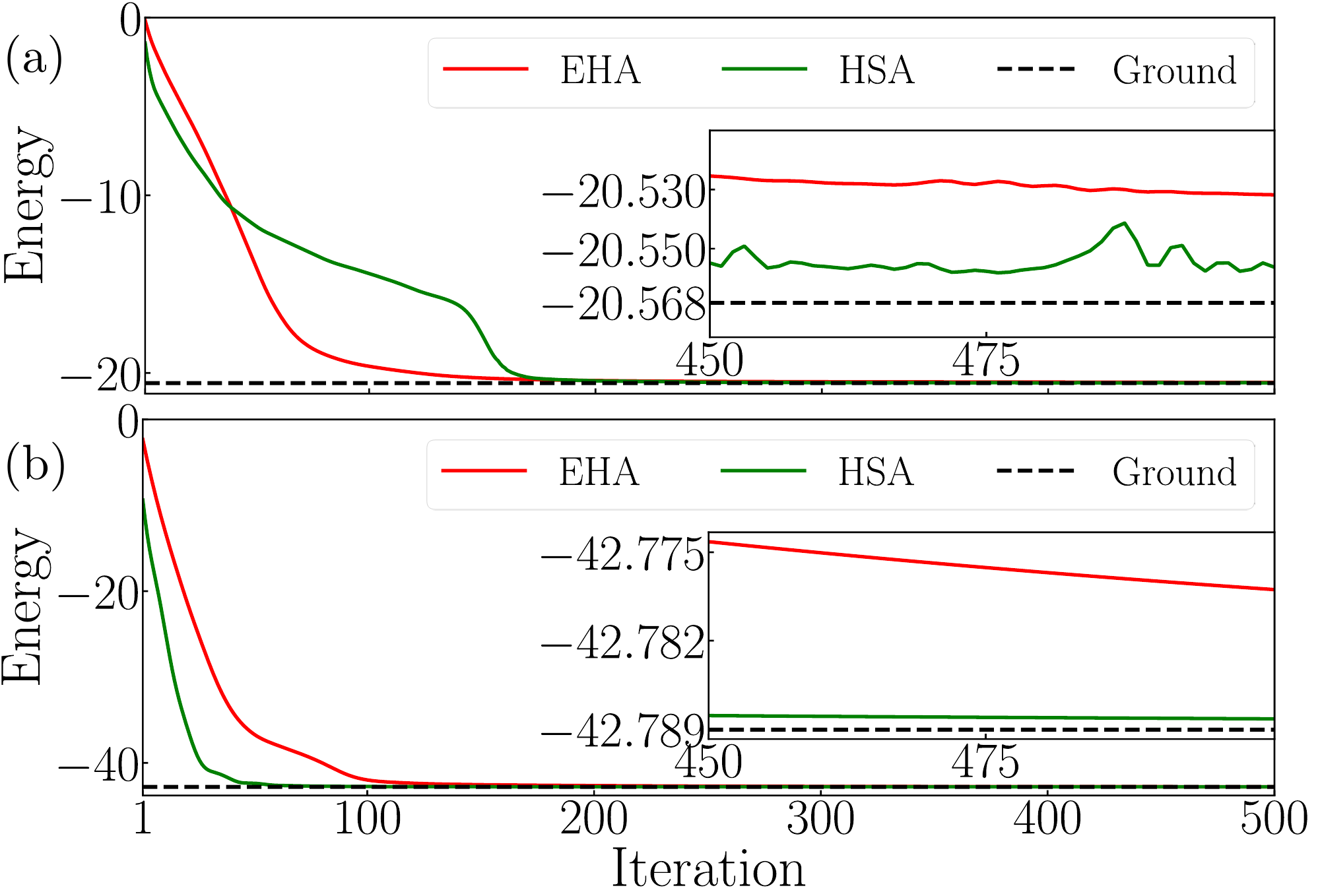}
	\caption{The best performance of EHA and HSA among~10 experiments for (a) HM in Eq.~(\ref{eq:XXZ}) and (b) TFIM in Eq.~(\ref{eq:TFIM}). The insets illustrate their fine differences.  HSA is slightly better than our EHA in obtaining a lower energy.}
	\label{fig:best}
\end{figure}

Recall that HSA is a problem-specific ansatz that is utilized to improve the performance of VQE. However,  from Fig.~\ref{fig:res_XXZ} and Fig.~\ref{fig:TFIM_cost}, it is clear that the sample variance of HSA is much larger than other ansatzes. This implies that its performance relies on different initial parameters. In fact, in the above experiments, it does not always provide a good approximation to the actual solution. However, it does converge to the solution approximately under some initializations. This is demonstrated in Fig.~\ref{fig:best}, where we plot the best performance of our EHA and HSA among 10 experiments for the HM in Eq.~(\ref{eq:XXZ}) and TFIM in Eq.~(\ref{eq:TFIM}). We find that from the perspective of the best performance, the problem-specific HSA is slightly better than our EHA.  Nevertheless,  it is important to note that our ansatz is problem-agnostic, and it is more robust to different realizations.

\begin{figure}[htpb]
	\centering
	\includegraphics[width=0.45\textwidth]{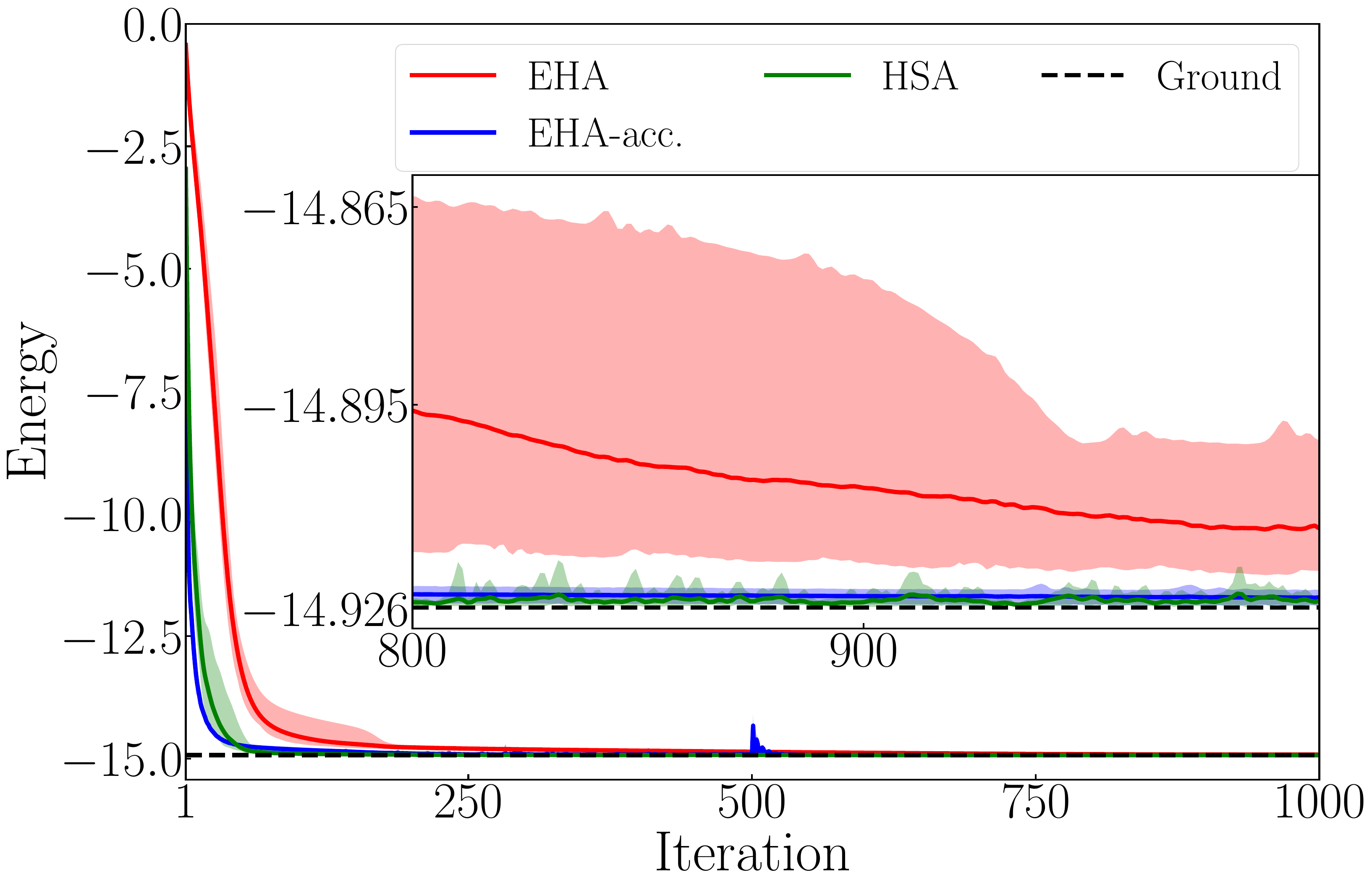}
	\caption{Comparison between our EHA and HSA for TFIM in  Eq.~(\ref{eq:TFIM1}). Although HSA outperforms EHA, the performance of EHA with acceleration is similar to that of HSA.}  
	\label{fig:TFIM1}
\end{figure}

We then consider the TFIM in  Eq.~(\ref{TFIM}) with $J_z=h_x=-1$, whose Hamiltonian reads
\begin{equation}
	\label{eq:TFIM1}
\begin{aligned}
H_{\mathrm{TFIM2}} = - \sum_{i=1}^{11} \sigma^z_i \sigma^z_{i+1} - \sum_{i=1}^{12}   \sigma^x_i.
\end{aligned}
\end{equation}
 Recall that for the Hamiltonian Eq.~(\ref{TFIM}), a quantum phase transition occurs at $J_z=h_x$, and at this critical point, the ground state is highly entangled and in a complex form~\cite{lyu2023symmetry}. We compare our $10$ blocks EHA and $20$ blocks HSA, and demonstrate the experimental results in Fig.~\ref{fig:TFIM1}. We find that for the TFIM in  Eq.~(\ref{eq:TFIM1}),  HSA converges to the actual ground state energy closer and faster than our EHA when using a step size of 0.01. To accelerate the convergence rate of EHA, we can increase its step size to 0.05 for the first 500 iterations and keep 0.01 for the rest. The accelerated version of EHA is referred to as EHA-acc. and its experimental results are also demonstrated in  Fig.~\ref{fig:TFIM1}. We find that the performance of  EHA-acc. is similar to that of HSA. 

\subsubsection{Quantum chemistry problems}

Now we compare our EHA with chemically inspired ansatzes UCCSD and GRSD by focusing on the $\mathrm{HF}$ molecule and $\mathrm{H}_3^{+}$ cation.  The quantum circuits for UCCSD and GRSD are much more complicated than our EHA. To be specific, from Fig.~\ref{fig:gate} and Fig.~\ref{fig:QERA}, the entangling gates in our EHA are all local CX gates that are applied to nearest-neighbor qubits. However, when realizing UCCSD, it has some non-local CX gates, while for GRSD, in addition to non-local CX gates, it has some T gates. We list the number of all CX gates, non-local CX gates and T gates of UCCSD and GRSD for the $\mathrm{HF}$ molecule and $\mathrm{H}_3^{+}$ cation in Table~\ref{tab:resource} (Appendix~\ref{app:chem}).
Since most current hardware platforms only allow nearest-neighbor connections~\cite{linke2017experimental,sanaei2019qubit}, and the use of T gates causes a very high cost~\cite{orts2023improving}, our EHA is much easier to be realized with current NISQ devices compared to UCCSD and GRSD. Thus, in the following we only focus on the performance of EHA, UCCSD and GRSD without assuming that they have roughly the same number of CX gates.

Recall that we  adopt \texttt{PennyLane} to simulate UCCSD and GRSD (referred to as ALLSD in \texttt{PennyLane}).  For both of them, we start with the Hartree-Fock state~\cite{google2020hartree}. For our EHA,  we  project the final output states to the feasible state space, which is spanned by the states considering all single and double excitations above the Hartree-Fock state. We show that our  EHA is superior to both of them in  robustly obtaining a lower energy.

\begin{figure}[htpb]
	\centering
	\includegraphics[width=0.45\textwidth]{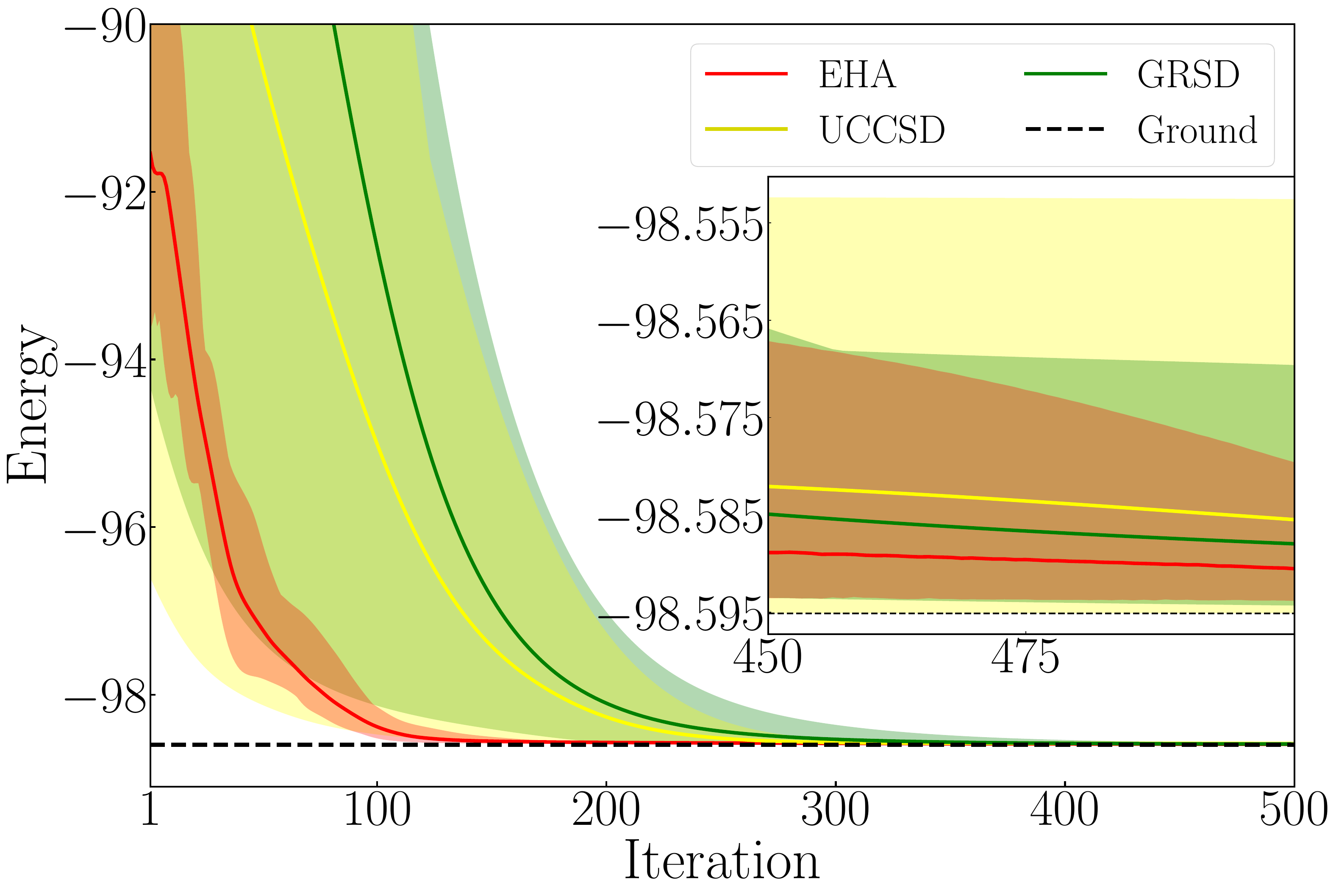}
	\caption{Comparison of our EHA with UCCSD and GRSD for $\mathrm{HF}$. The black dashed line denotes the ground energy of the $\mathrm{HF}$ Hamiltonian  obtained by second quantization. Our EHA can robustly achieve a lower average energy as compared to UCCSD and GRSD. }
	\label{fig:HF}
\end{figure}

As for the $\mathrm{HF}$ molecule, its bond length is set to be $1.1$. Our EHA consists of 12 qubits and has 18 blocks.  We demonstrate the experimental results in Fig.~\ref{fig:HF}.  Here, the black dashed baseline denotes the ground energy of the $\mathrm{HF}$ Hamiltonian  obtained by second quantization.
We find that our EHA can achieve a lower energy on average and its performance is more robust with respect to different realizations compared to UCCSD and GRSD. While for the best performance among  10 realizations, UCCSD and GRSD are slightly superior to our EHA.

\begin{figure}[htpb]
	\centering
	\includegraphics[width=0.45\textwidth]{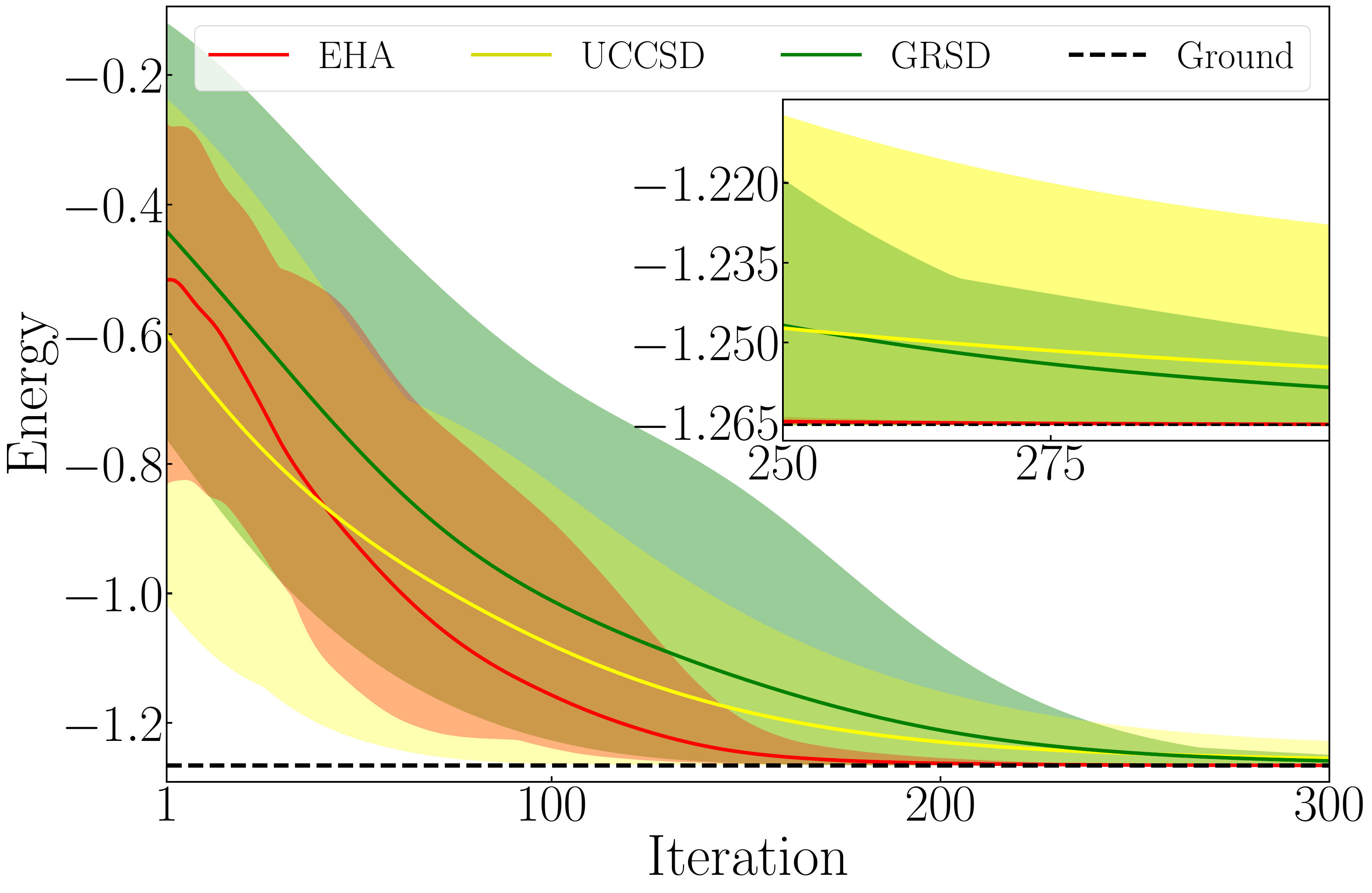}
	\caption{Comparison of our EHA with UCCSD and GRSD for $\mathrm{H}_{3}^{+}$. The black dashed line denotes the ground energy of the second quantization Hamiltonian of  $\mathrm{H}_3^{+}$. Our EHA can always achieve the ground energy. }
	\label{fig:H3}
\end{figure}

As for the $\mathrm{H}_3^{+}$ cation,  we need to modify the cost function Eq.~(\ref{eq:Cost})  by adding a penalty term for our EHA. The reason is as follows. On the one hand the second quantization Hamiltonian of $\mathrm{H}_3^{+}$ depends only on the spin-orbital basis, and is independent of the number of electrons~\cite{helgaker2013molecular}. On the other hand, in contrast to UCCSD and GRSD, our EHA does not preserve particle numbers. Therefore, if there is no constraint, under EHA the trial states will converge to the ground state of $\mathrm{H}_3$ having 3 electrons, while $\mathrm{H}_3^{+}$  has only  $2$ electrons. To address this issue, we can simply add a constraint on the number of  electrons by means of penalty function. Specifically, we consider the cost function as
\begin{equation}
\begin{aligned}
	\label{eq:pen}
C(\theta) = \bra{\psi(\theta)}H\ket{\psi(\theta)}  + \beta [\bra{\psi(\theta)} N \ket{\psi(\theta)} -2  ]^2.
\end{aligned}
\end{equation}
Here, $H$ is the Hamiltonian of $\mathrm{H}_{3}^{+}$, $\beta$ is the penalty hyperparameter, and  $N = \sum_{\alpha}^{ } \hat{c}_{\alpha}^{\dagger} \hat{c}_{\alpha} $ denotes  the number operator with  $\hat{c}^{\dagger}_{\alpha}$ and $\hat{c}_{\alpha}$ being  the particle creation and annihilation operators, respectively, and the index $\alpha$ running over the basis of single-particle. Here, for $\mathrm{H}_3^{+}$, the bond length is set to be $1.1$. Our EHA consists of 6 qubits and has 9 blocks, and the hyperparameter $\beta = 10$.  The experimental results are shown in Fig.~\ref{fig:H3}. We find that our EHA is much better than UCCSD and GRSD, as EHA can always achieve the ground energy for different realizations.

\subsection{Ability to quickly adjust entanglment}
\label{sec:ent}
Recall that to solve the ground state problem, it is imperative  to generate a quantum state with the matched entanglement.
In this subsection, we demonstrate that because of the variational entangler design in our EHA, it can rapidly adjust the entanglement to the required amount during the training process.

For an $N$-qubit system state $\rho$, denote by $\rho_i$ the state of the $i$-th qubit by taking partial trace over all the other qubits, namely, $\rho_i = \operatorname{Tr}_{\overline{i}}[\rho]$, where $\overline{i}$ denotes the subsystem excluding the $i$-th qubit. Here,  we adopt the average von Neumann entropy
\begin{equation*}
\begin{aligned}
S  \triangleq -\frac{1}{N} \sum_{i=1}^{N} \operatorname{Tr} \left[ \rho_i \log(\rho_i) \right]
\end{aligned}
\end{equation*}
to be the figure of merit for evaluating the entanglement of quantum state $\rho$.

For models already investigated in Subsection~\ref{sec:eff}, we now investigate their entanglement transitions  of the generated trial states under different ansatzes during the optimization. In the following figures of this subsection, the black dashed line denotes the entanglement of the ground state, and the other lines denote the respective average value of the entanglement over 10 realizations under different ansatzes.

\begin{figure}[htpb]
	\centering
	\includegraphics[width=0.45\textwidth]{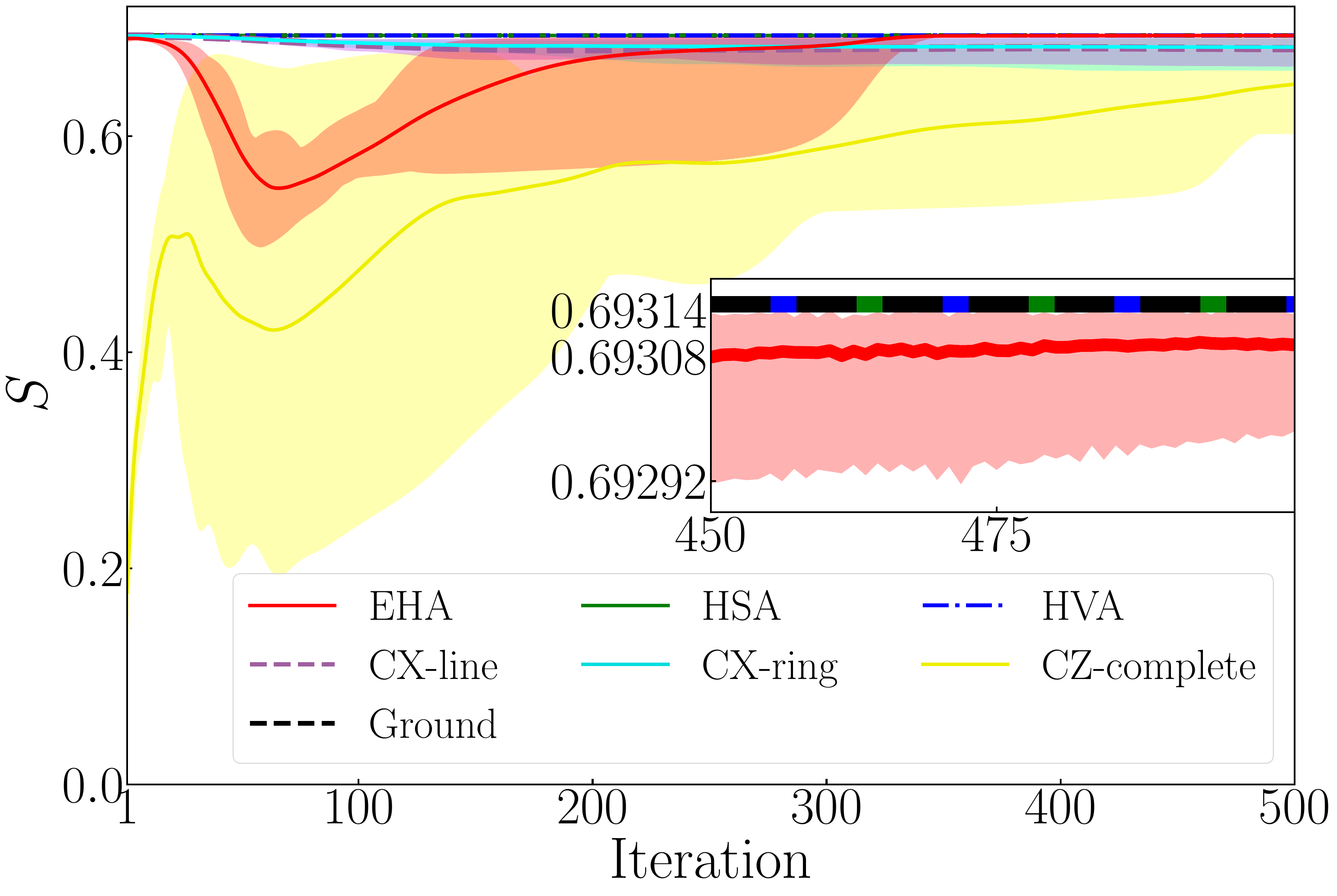}
	\caption{Entanglement transitions under different ansatzes for HM in Eq.~(\ref{eq:XXZ}).  Our EHA can quickly adjust the entanglement to the desired amount.}
	\label{fig:XXZ_ent}
\end{figure}

For the $12$-qubit HM in Eq.~(\ref{eq:XXZ}), we illustrate the entanglement transitions (corresponding to Fig.~\ref{fig:res_XXZ})  in  Fig.~\ref{fig:XXZ_ent}. We find that the ground state has a large amount of entanglement. Under  HVA and HSA, the entanglement of the generated quantum states stays at the same value as the desired amount during the optimization. This validates that for the problem-specific HVA and HSA, they only explore specific spaces of the unitaries during training. This is the main reason that HVA and HSA usually outperform problem-agnostic HEAs. However, as we mentioned, the performance of HSA and HVA depends heavily on the initial reference states (see Appendix~\ref{app:HSA}), which are often hard to determine.  For problem-agnostic HEAs,  it is clear that our EHA can rapidly adjust the entanglement to the desired level, while  the other HEAs cannot. This is owing to the variational entangler design in our EHA.

\begin{figure}[htpb]
	\centering
	\includegraphics[width=0.45\textwidth]{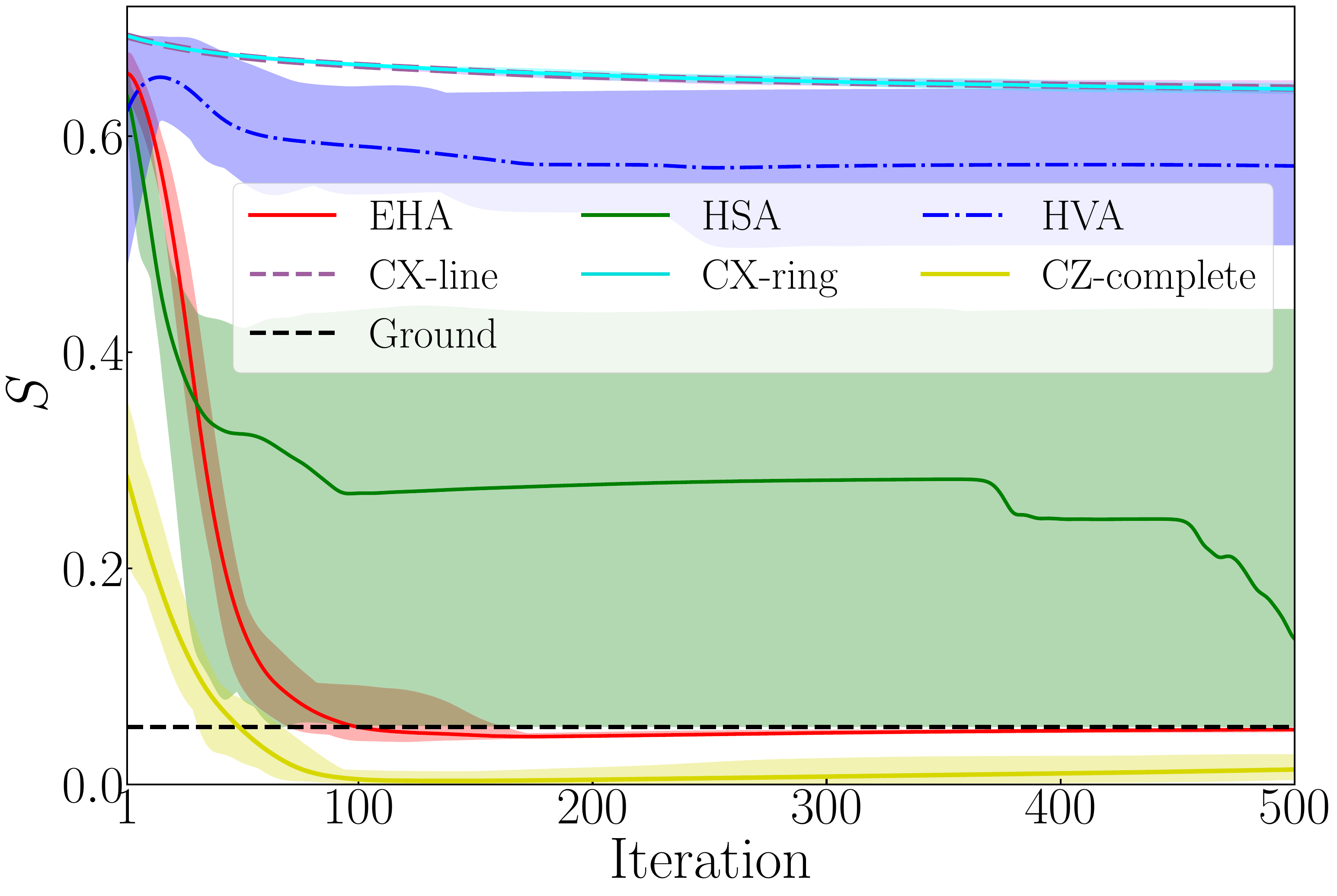}
	\caption{Entanglement transitions under different ansatzes for TFIM in Eq.~(\ref{eq:TFIM}).~Our EHA is much better than other ansatzes in adjusting the entanglement to the required amount.}
	\label{fig:TFIM_ent}
\end{figure}

For the $12$-qubit TFIM in Eq.~(\ref{eq:TFIM}), we illustrate the entanglement transitions (corresponding to Fig.~\ref{fig:TFIM_cost}) in Fig.~\ref{fig:TFIM_ent}.  We find that in this case the entanglement of the ground state is low. Under our EHA, the excess entanglement can  be quickly removed and adjusted to the desired amount,  which is hard to do with other ansatzes. The robustness of the entanglement transitions under different realizations is particularly poor for  HSA.

\begin{figure}[htpb]
	\centering
	\includegraphics[width=0.45\textwidth]{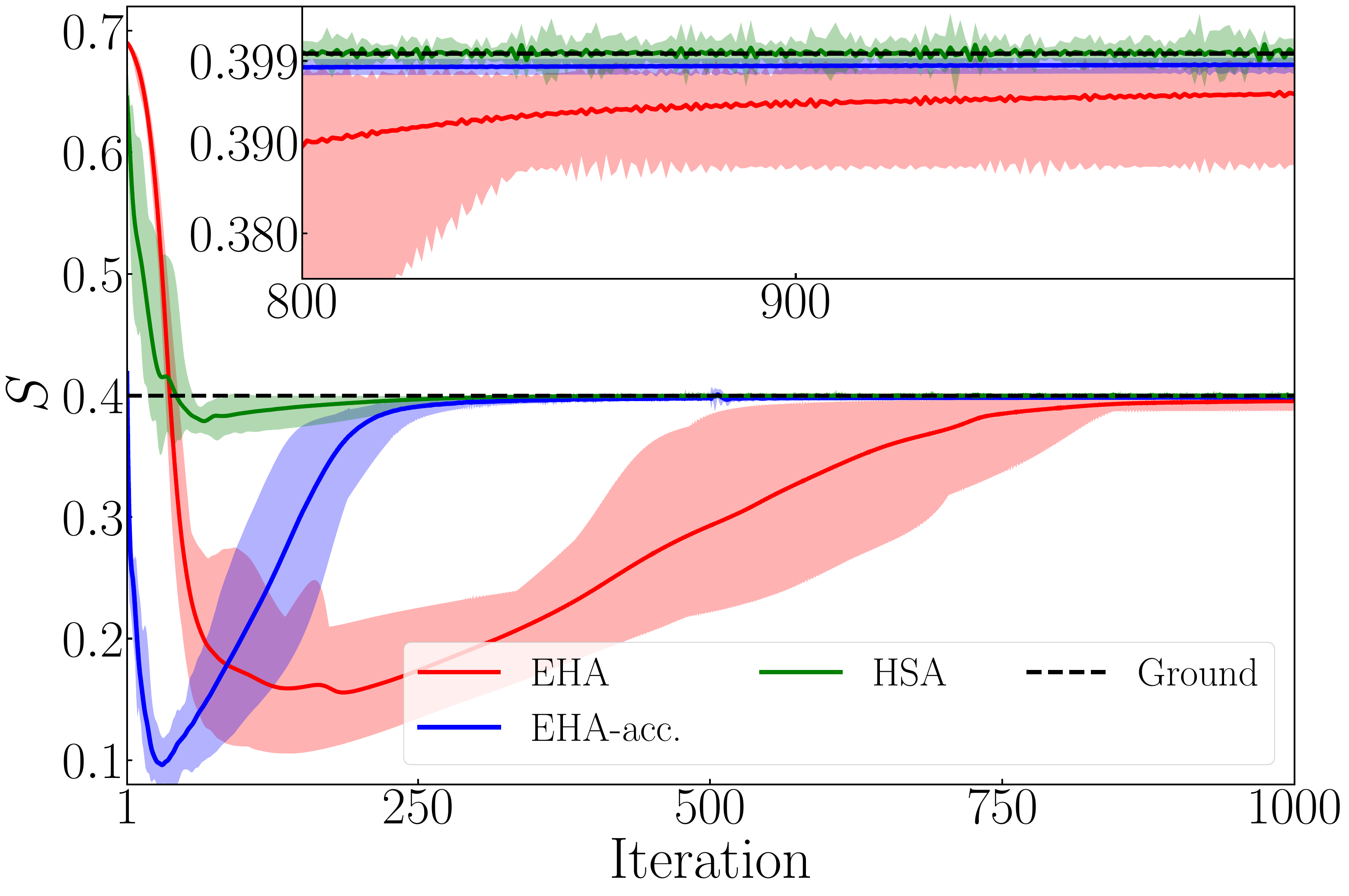}
	\caption{Entanglement transitions under HSA, EHA and EHA-acc. for TFIM in Eq.~(\ref{eq:TFIM1}).  }
	\label{fig:TFIM1_ent}
\end{figure}

As for the TFIM in Eq.~(\ref{eq:TFIM1}), we demonstrate the entanglement transitions under EHA and HSA in Fig.~\ref{fig:TFIM1_ent}. From Fig.~\ref{fig:TFIM1} and Fig.~\ref{fig:TFIM1_ent}, we find that in this case
HSA is superior to our EHA, as it can adjust the  entanglement to the required level in a much quicker way.  However,  we note that the accelerated version of EHA can also tune the entanglement to the required level quickly.

\begin{figure}[htpb]
	\centering
	\includegraphics[width=0.45\textwidth]{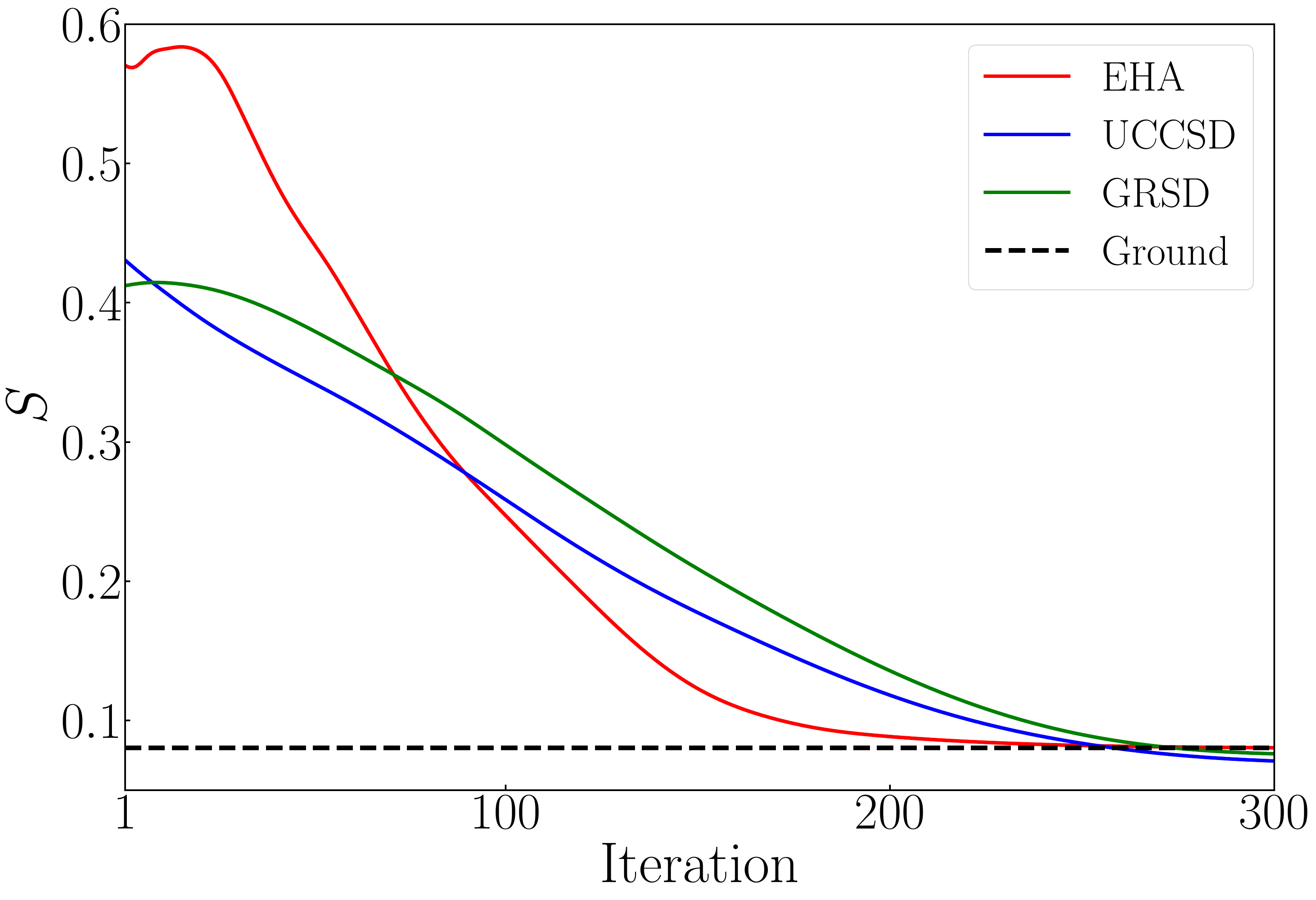}
	\caption{Entanglement transitions under EHA, UCCSD and GRSD for $\mathrm{H}_3^{+}$.  }
	\label{fig:H3_ent}
\end{figure}

For the $\mathrm{H}_3^{+}$ cation, the entanglement transitions under EHA, UCCSD and GRSD (corresponding to Fig.~\ref{fig:H3}) are shown in Fig.~\ref{fig:H3_ent}. Here, we only depict the average value of the entanglement over 10 realizations.  We find that as compared with UCCSD and GRSD, EHA can quickly adjust the entanglement to the desired level.

\subsection{The impact of initial reference states}

In this subsection, we consider the impact of initial reference states on our EHA. Since the reference state in quantum chemistry problems is typically chosen as Hartree-Fock state, we only concern quantum many-body systems. Moreover, for problem-specific ansatzes like HSA and HVA, the selection of initial reference states has strict requirements. Taking the HSA as an example, starting from an initial state that violates the symmetry of the Hamiltonian will automatically create a bias against HSA resulting in poor performance. This has been illustrated in Fig.~\ref{fig:HSA_init} (Appendix~\ref{app:HSA}).
Therefore, we analyze the impact of different initial reference states on ansatzes only by comparing the performance of EHA and HEA. Note that in the above experiments, among the three considered HEAs, the CZ-complete performs best. We now compare EHA with CZ-complete under different initial reference stats.

\begin{figure}[htpb]
	\centering
	\includegraphics[width=0.45\textwidth]{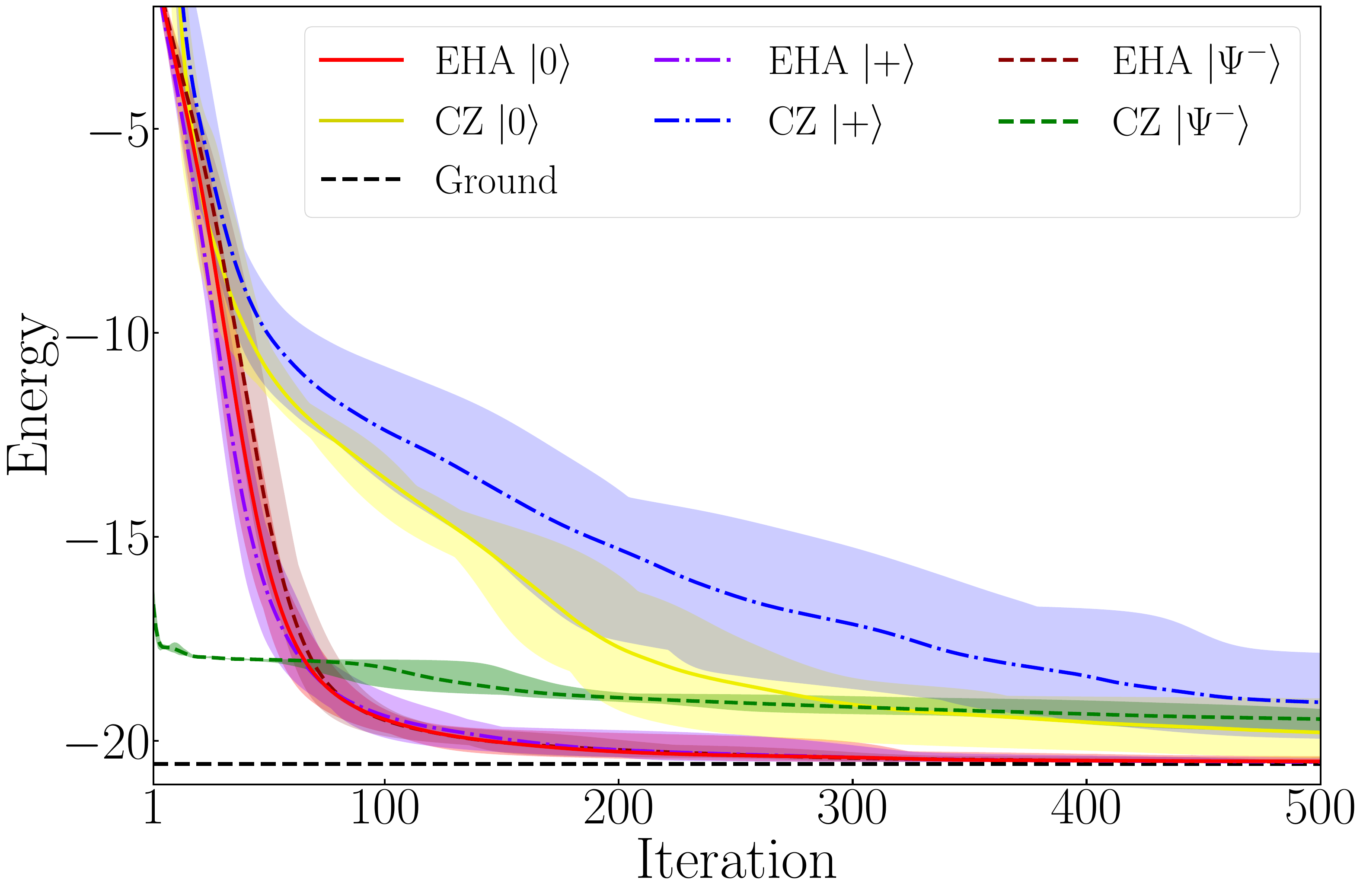}
	\caption{Comparison between our EHA and CZ-complete under different initial reference states for HM in Eq.~(\ref{eq:XXZ}). The performance of our EHA is more robust than CZ-complete.} 
	\label{fig:XXZ_cost_init}
\end{figure}

\begin{figure}[htpb]
	\centering
	\includegraphics[width=0.45\textwidth]{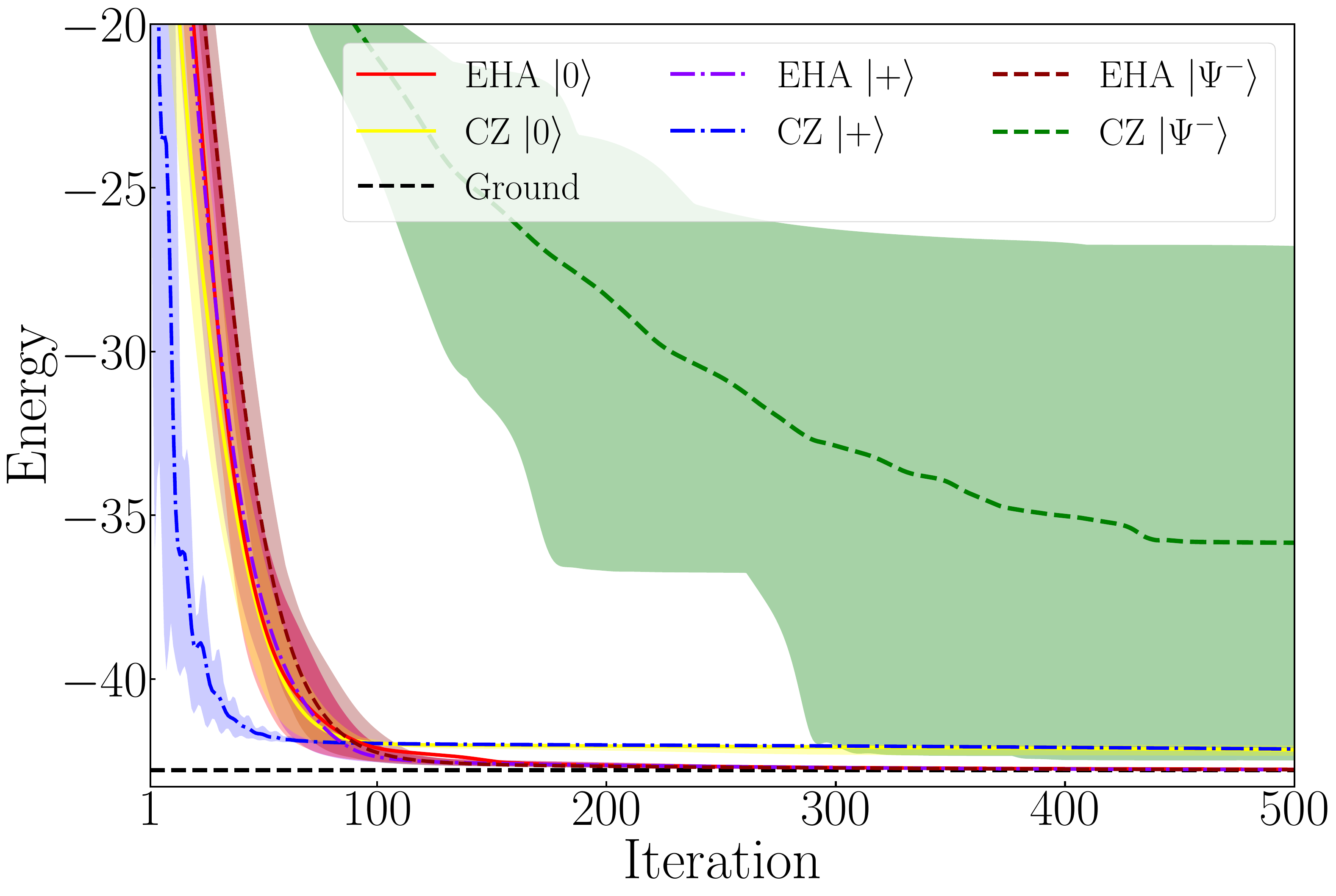}
	\caption{Comparison between our EHA and CZ-complete under different initial reference states for TFIM in Eq.~(\ref{eq:TFIM}). The performance of our EHA is more robust than CZ-complete. }
	\label{fig:TFIM_cost_init}
\end{figure}

\begin{figure}[htpb]
	\centering
	\includegraphics[width=0.44\textwidth]{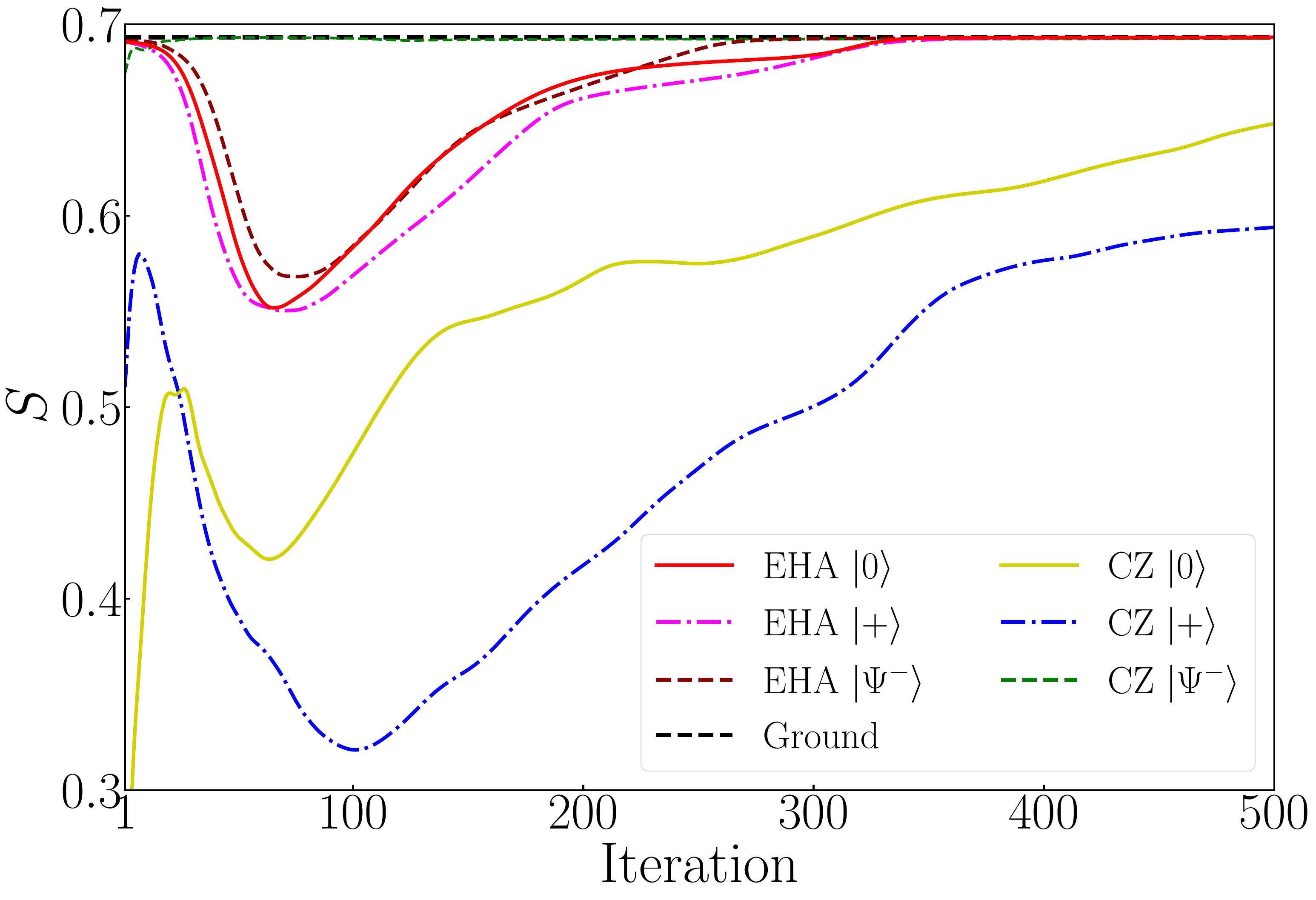}
	\caption{Entanglement transitions of EHA and CZ-complete under different initial reference states for HM in Eq.~(\ref{eq:XXZ}). The black dashed line denotes the entanglement of the ground state. Other lines denote the respective average value of the entanglement over 10 realizations. The entanglement transitions of our EHA are more robust than CZ-complete under different initial states. }
	\label{fig:XXZ_ent_init}
\end{figure}

We consider three typical initial reference states: ${\otimes^{12}}\ket{0} $, $ {\otimes^{12}}\ket{+} $ and ${\otimes^{6}} \ket{\Psi^{-}} $. The experimental results for HM in Eq.~(\ref{eq:XXZ})  and TFIM in Eq.~(\ref{eq:TFIM}) are demonstrated in Fig.~\ref{fig:XXZ_cost_init} and Fig.~\ref{fig:TFIM_cost_init}, respectively. We find that for both HM and  TFIM, our EHA is superior to CZ-complete, and the performance of EHA is more robust under different initial reference states. The phenomenon can be explained by checking the corresponding  entanglement transitions. As an illustration, we demonstrate the entanglement transitions of HM (corresponding to Fig.~\ref{fig:XXZ_cost_init}) in Fig.~\ref{fig:XXZ_ent_init}.
It is clear that compared to CZ-complete, our EHA can robustly and quickly adjust the entanglement to the desired level under different initial states. This is due to the variational design in EHA, which also makes EHA a strong candidate for quantum computing.  

\subsection{Avoiding BP with reduced-domain initialization}

In this subsection, we consider how to train our EHA with relatively large blocks.

As illustrated in Fig.~\ref{fig:gate} and Fig.~\ref{fig:QERA}, our EHA utilizes a total number of $6(N-1)$ CX gates per block. This makes the expressibility of EHA grow quickly as the number of blocks $L$ increases. However, it is well known that too expressive ansatz will result in the phenomenon of BP, making the training extremely difficult. Therefore, our EHA may suffer from BP when $L$ is large.

There have been many ways to mitigate the impact of BP~\cite{sack2022avoiding,friedrich2022avoiding,mele2022avoiding,wang2023trainability,zhang2022escaping,skolik2021layerwise}. Here, we adopt the reduced-domain method introduced in~\cite{wang2023trainability}. It was stated that to balance the conflict between trainablity and expressibility of PQCs,  the domain of each variational parameter should be reduced in proportion to $\frac{1}{\sqrt{L}}$, where $L$ denotes the depth of  PQC \cite{wang2023trainability}.  We demonstrate that the reduced-domain  initialization~\cite{wang2023trainability} can help train our EHA with relatively large blocks.

\begin{figure}[htpb]
	\centering
	\includegraphics[width=0.45\textwidth]{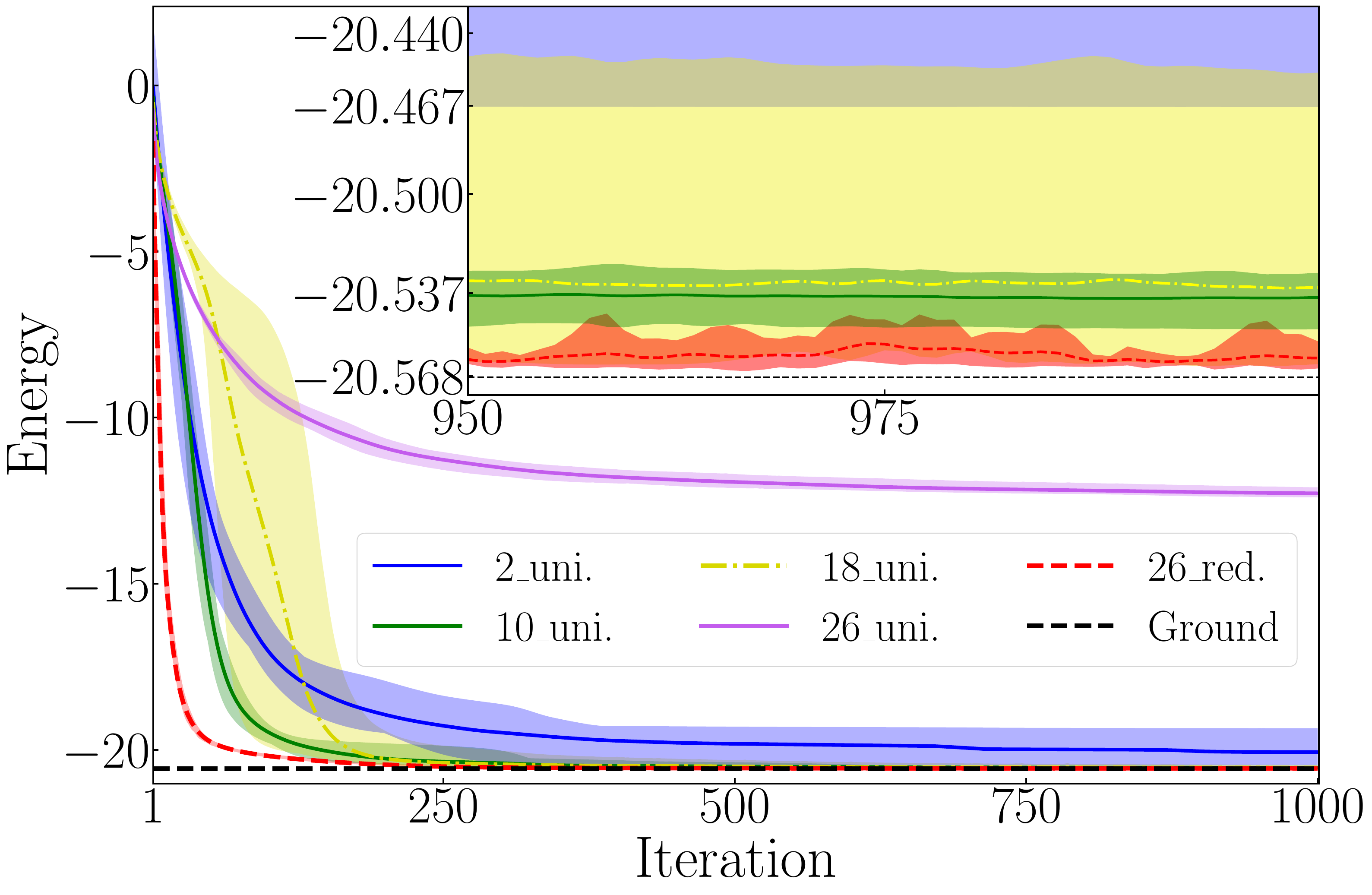}
	\caption{Performance of EHA under different numbers of blocks and initializations for HM in Eq.~(\ref{eq:XXZ}).  With the reduced-domain initialization, the 26-block circuit can generate quantum states closer to the ground state in a faster way compared to other schemes. }
	\label{fig:over}
\end{figure}

We focus on the $12$-qubit HM in Eq.~(\ref{eq:XXZ}), and consider four different numbers of blocks, namely, $L=2,10,18$ and $26$.
The experimental settings are the same as those in Subsection~\ref{sec:eff}, except  for the initialization.

We first draw the variational parameters according to the uniform distribution $\mathcal{U}[-\pi,\pi]$ as we have done in Subsection~\ref{sec:eff}. The numerical results are illustrated in Fig~\ref{fig:over}. We find that the  $10$-block EHA has the best performance among the four considered cases, as it robustly converges to a lower energy. As compared with the 10-block EHA, the 2-block EHA is less expressible and cannot obtain better performance, while the 18-block EHA is more expressible but difficult to be trained~\cite{holmes2022connecting}, which is indicated by the slower convergence rate and larger sample variance. The 26-block EHA is too expressive to be trained. To address the training issue for large blocks of EHA, we leverage the reduced-domain initialization method~\cite{wang2023trainability}. Specifically, each variational parameter is drawn according to a reduced uniform distribution $\mathcal{U}[\frac{\pi}{2}-\frac{1}{\sqrt{L}},\frac{\pi}{2}+\frac{1}{\sqrt{L}}]$.  The performance of the $26$-block reduced version of EHA is plotted in Fig.~\ref{fig:over}. We find that with the reduced uniform initialization, the 26-block EHA has a much faster convergence rate, higher level of accuracy and robustness with respect to different realizations, as compared with other cases.

\subsection{Performance Testing}

In this subsection, we focus on the performance of our EHA on additional quantum chemistry and quantum many-body models.

We first compare our EHA with an ansatz termed adaptive derivative-assembled pseudo-Trotter ansatz variational quantum eigensolver (ADAPT-VQE), which was proposed in~\cite{grimsley2019adaptive} for molecular simulations.  The key idea of ADAPT-VQE is to systematically grow the ansatz by adding fermionic operators one at a time to maximumly recover the correlation energy at each step.
The initial reference state of ADAPT-VQE is also the Hartree-Fock state, and the operator pool includes all single  and double excitation operators acting on the reference state without flipping the spin of the excited particles. We list the number of all CX gates, non-local CX gates and T gates of ADAPT-VQE for different molecules in Table~\ref{tab:resource} (Appendix~\ref{app:chem}).  It has been shown that  ADAPT-VQE outperforms UCC in terms of both circuit depth and chemical accuracy~\cite{grimsley2019adaptive}.

For all molecules, the  ADAPT-VQE results in this paper are obtained by \texttt{PennyLane}, with the termination condition being gradient less than $10^{-3}$. While for our EHA, we find that for all experiments, the gradient is never below $10^{-3}$ before the end of training. This implies that the gradient of EHA is larger than ADAPT-VQE before the end of the optimization. Moreover, for all molecules, we have checked that the spin and particle number of the solutions obtained by both ADAPT-VQE and EHA are the same as the actual states.

\begin{figure}[htpb]
	\centering
	\includegraphics[width=0.45\textwidth]{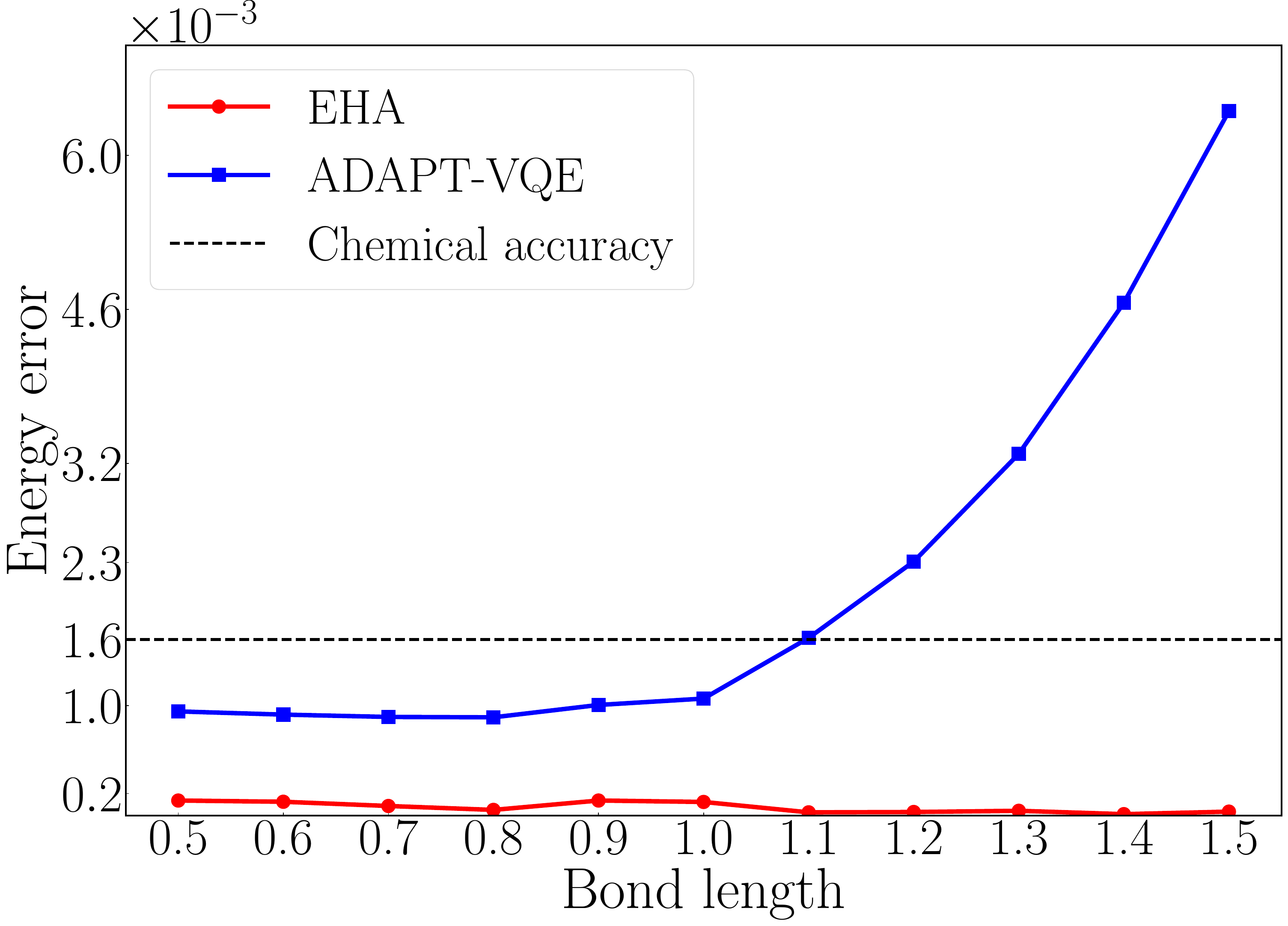}
	\caption{Ground energy error of   EHA  and ADAPT-VQE for  $\mathrm{H}_5$. The energy error obtained by ADAPT-VQE exceeds the chemical accuracy and grows quickly  when the bond length is larger than 1.1. While for our EHA, the energy error stays nearly at the same value as the bond length increases, and is  about $1/8$  of the chemical accuracy. }
	\label{fig:H5}
\end{figure}

\begin{table*}
	\begin{ruledtabular}
	\begin{tabular}{cccccccc}
   Model &Qubits & Blocks  & Ground energy & ADAPT-VQE energy & EHA-best &  EHA-mean & EHA-STD\\
	\hline
	$\mathrm{BeH}_2$(1.1) &14   & 32       & -15.5496      & -15.5490   & -15.5490          & -15.5490           & 0.0000\\
	LiH(1.11)              &12  & 16       & -7.8288       & -7.8282    & -7.8286           & -7.8286            & 0.0000 \\
	HF(1.1)                &12  & 18       &-98.5951     &-98.5942     & -98.5948          &-98.5947                 &0.0000 \\
	\end{tabular}
	\end{ruledtabular}
	\caption{\label{tab:QERA_abs0}Performance of EHA and ADAPT-VQE for different molecules.
}
\end{table*}

\begin{table*}
	\begin{ruledtabular}
	\begin{tabular}{ccccccccc}
    Model  & Blocks  & Ground energy  & Energy-best & Energy-mean & Energy-STD & Fidelity-best & Fidelity-mean & Fidelity-STD\\
	\hline
	HM(8)          & 14       &-13.4997               &-13.4994           & -13.4993           & 0.0001   & 1.0000 & 1.0000 & 0.0000 \\
	HM(12)         & 28       &-20.5684               &-20.5679           & -20.5675           & 0.0002   & 1.0000 & 0.9999 & 0.0000 \\
	HM(16)         & 42       &-27.6469               &-27.6461           & -27.6459           & 0.0002   & 0.9999 & 0.9999 & 0.0000 \\
	TFIM1(8)      & 6        & -20.5018              & -20.5018          & -20.5015           & 0.0004   & 1.0000 & 1.0000 & 0.0000 \\
	TFIM1(12)     & 6        & -42.7890              & -42.7889          & -42.7880           & 0.0010   & 1.0000 & 1.0000 & 0.0000 \\
	TFIM1(16)     & 8        & -57.0763              & -57.0760          & -57.0759           & 0.0002   & 1.0000 & 1.0000 & 0.0000 \\
	TFIM2(8)      & 8        & -9.8380               & -9.8378           & -9.8376            & 0.0002   & 1.0000 & 0.9999 & 0.0000 \\
	TFIM2(12)     & 12       & -14.9260              & -14.9257          & -14.9255           & 0.0002   & 1.0000 & 0.9999 & 0.0000 \\
	TFIM2(16)     & 20       & -20.0164              & -20.0160          & -20.0156           & 0.0003   & 0.9999 & 0.9999 & 0.0000 \\
	BHM(8)         & 12       & 0.0000                & 0.0001            &0.0002              & 0.0001   & 1.0000 & 1.0000 & 0.0000 \\
	BHM(16)        & 20       & 0.0000                & 0.0001            &0.0002              & 0.0001   & 1.0000 & 1.0000 & 0.0000 \\
	\end{tabular}
	\end{ruledtabular}
	\caption{\label{tab:QERA_abs}Performance of EHA for different quantum many-body systems.
}
\end{table*}

We first consider solving the ground state of $\mathrm{H}_5$  at different bond lengths. Here, we consider the target ground state with $s_z=\frac{1}{2}$.  Our EHA  consists of $10$ qubits,  $L = 38$ blocks, and the initial parameters are drawn from a reduced uniform distribution  $\mathcal{U}[\frac{\pi}{2} - \frac{1}{\sqrt{L}},\frac{\pi}{2} + \frac{1}{\sqrt{L}}]$. To guarantee the obtained state having $s_z=\frac{1}{2}$, similar to what we have done in dealing with $\mathrm{H}^+_3$, we add an $S_z$ penalty term (with the hyperparameter being 100) in the cost function in our EHA. We perform 2000 iterations, with the step size being 0.1 for the first 500 iterations, and 0.001 for the last 1500 iterations. We take the minimum energy during the training process as the approximate solution of the ground energy.  We illustrate the ground energy error in Fig.~\ref{fig:H5}. We find that when the bond length is larger than 1.1, the ground energy error obtained by ADAPT-VQE exceeds the chemical accuracy (1.6$\times$ $10^{-3}$)~\cite{eyring1935activated} and grows quickly as the bond length increases. While for our EHA, the energy error remains nearly constant as the bond length increases, and is lower than ADAPT-VQE. Specifically, the ground energy error can maintain about $1/8$  of the chemical accuracy for bond length ranging from 0.5 to 1.5.

We now apply our EHA and ADAPT-VQE to other molecules. For our EHA we perform at most $3000$ iterations, and for different molecules the step size schedules are illustrated in Table~\ref{tab:iter_schedule} (Appendix~\ref{app:step}).  The results are demonstrated in Table~\ref{tab:QERA_abs0}.  The first column lists the molecules with the bond length in parentheses, the second column shows the  number of qubits needed after second quantization of the molecules, and the third column lists the blocks of our EHA. We also list the best value, mean value and the standard deviation (STD) of the energy obtained by our EHA in Table~\ref{tab:QERA_abs0}. It is clear that our EHA outperforms ADPT-VQE and has less than $5\times 10^{-5}$ standard deviation.

We further apply our EHA on other quantum many-body systems, where their corresponding step size schedules are illustrated in Table~\ref{tab:iter_schedule} (Appendix~\ref{app:step}). We demonstrate the results in Table~\ref{tab:QERA_abs}.  The first column lists the models with the number of qubits in parentheses, where the HM, TFIM1 and TFIM2 are described by Eq.~(\ref{eq:XXZ}), Eq.~(\ref{eq:TFIM}) and Eq.~(\ref{eq:TFIM1}) respectively.
Here, the BHM denotes the Bose-Hubbard model (BHM)~\cite{fisher1989boson} in a chain lattice, whose Hamiltonian reads
\begin{equation}\label{BHM}
\begin{aligned}
H =-\left(\sum_{i=1}^{N-1} \hat{b}_{i}^{\dagger} \hat{b}_{i+1}+\text{h.c.}\right)+ 7 \sum_{i=1}^N \hat{n}_{i} (\hat{n}_{i}-1),
\end{aligned}
\end{equation}
where $\hat{b}_i^{\dagger}$ and $\hat{b}_i$ denote the Bosonic creation and annihilation operators on site $i$, respectively,   and  $\hat{n}_i = \hat{b}_i^{\dagger} \hat{b}_i$ denotes the number operator of site $i$.  For BHM, we can  use binary Bosonic mapping~\cite{huang2021qubitization} to transform the Hamiltonian in Eq.~(\ref{BHM}) into the form of  a linear combination of Pauli strings, which is done  by utilizing  \texttt{PennyLane}  in this paper.
The number of blocks in our EHA is listed in the second column. In Table~\ref{tab:QERA_abs}, we list the best value,  mean value and the standard deviation of the energy  and  fidelity obtained by our EHA. We find that for all cases, our EHA can attain the ground energy with a very high level of accuracy. In the best case, the ground energy error is less than $1 \times 10^{-3}$, and on average the ground energy error is less than the chemical accuracy of $1.6 \times 10^{-3}$. Moreover, the standard deviation is very small, which is at most $1 \times 10^{-3}$. The fidelity with the ground state is no smaller than $99.99\%$ on average, and the standard deviation is lower than $5 \times 10^{-5}$.

\section{Conclusion}
\label{sec:conclusion}

In this paper, we propose an efficient  hardware-efficient ansatz, EHA, for eigensolvers, in which the entanglers are designed to be variational rather than fixed. This entanglement-variational design allows the circuit to rapidly adjust the entanglement of the generated trial states to the required amount, and in turn greatly enhance the performance. We have demonstrated that our EHA can find approximate solutions for  eigensolvers with a very  high level of accuracy, and its performance is  robust to choices of  initial reference states and different realizations. We believe that our EHA is particularly suitable for NISQ era and will generate wide impact on developing algorithms with potential quantum advantages for various practical applications.

\bibliography{main}

\onecolumngrid

\pagebreak

\appendix

\section{CIRCUITS  OF PROBLEM-SPECIFIC ANSATZES}\label{AA}

For completeness, we illustrate the HVA and HSA circuits for HM  in Eq.~(\ref{XXZ}) and TFIM in Eq.~(\ref{TFIMequal}).

\begin{figure}[htpb]
	\centering
	\includegraphics[width=0.77\textwidth]{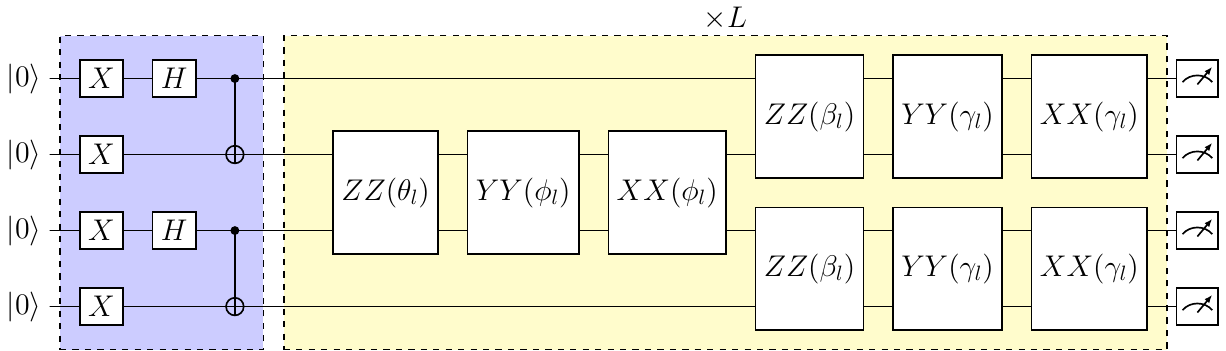}
	\caption{The HVA circuit for  HM in Eq.~(\ref{XXZ}). In HVA, the parameters of the entanglers are correlated.}
	\label{fig:HVA_XXZ}
\end{figure}

\begin{figure}[htpb]
	\centering
	\includegraphics[width=0.75\textwidth]{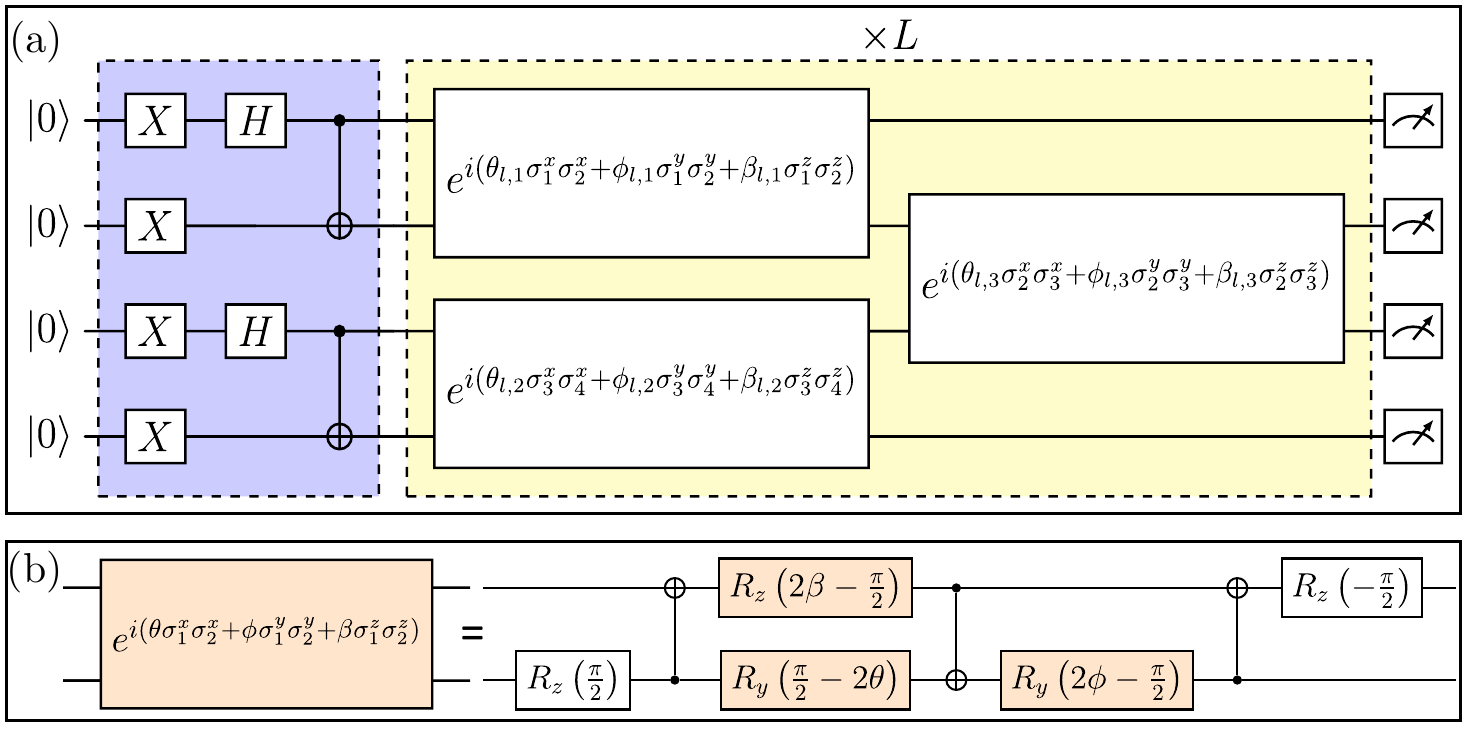}
	\caption{The HSA circuit for HM in Eq.~(\ref{XXZ}). (a) The architecture of HSA. (b) Circuit for realizing the entangler $e^{i(\theta \sigma^x_1 \sigma^x_2 + \phi\sigma^y_1 \sigma^y_2 + \beta\sigma^z_1 \sigma^z_2) }$ used in HSA.}
	\label{fig:HSA_XXZ}
\end{figure}

\begin{figure}[htpb]
	\centering
	\includegraphics[width=0.89\textwidth]{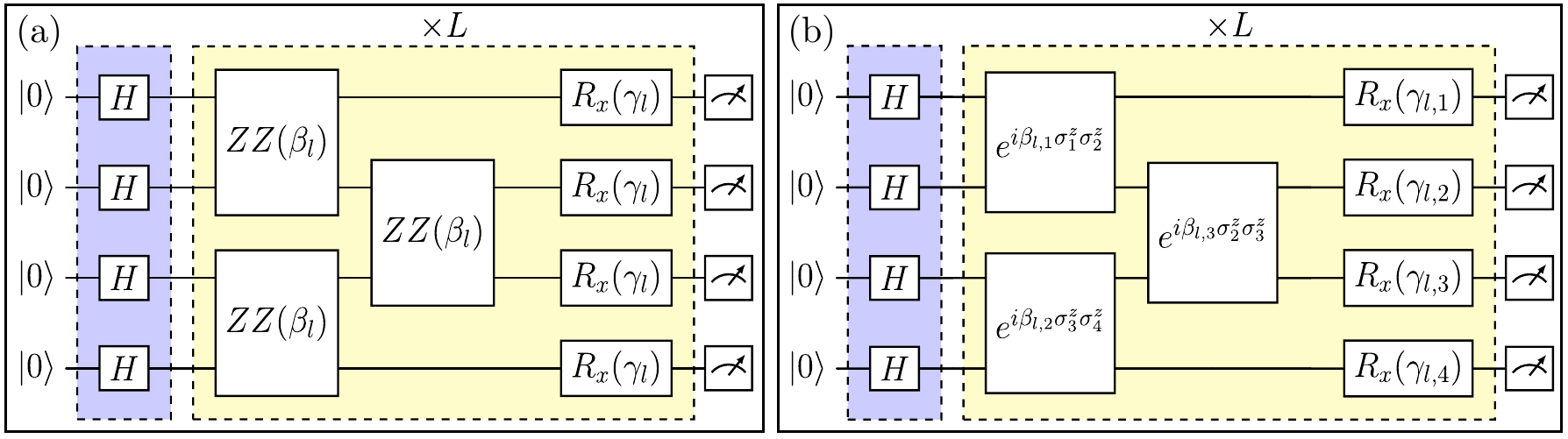}
	\caption{The HVA and HSA circuits for TFIM in Eq.~(\ref{TFIMequal}). (a) The HVA circuit. In each block, there are only two variational parameters, one for the single qubit rotational gates and the other for the entanglers. (b) The HSA circuit. The gates have different parameters in general. The entangler $e^{i\beta \sigma^z_1 \sigma^z_2}$ can be realized by setting $\theta = 0$ and $\phi = 0$ in Fig.~\ref{fig:HSA_XXZ}(b).}
	\label{fig:TFIM_HVA_HSA}
\end{figure}

\section{LIMITED MEASUREMENT SHOTS}
\label{app:shots}
\begin{figure}[H]
	\centering
	\includegraphics[width=0.95\textwidth]{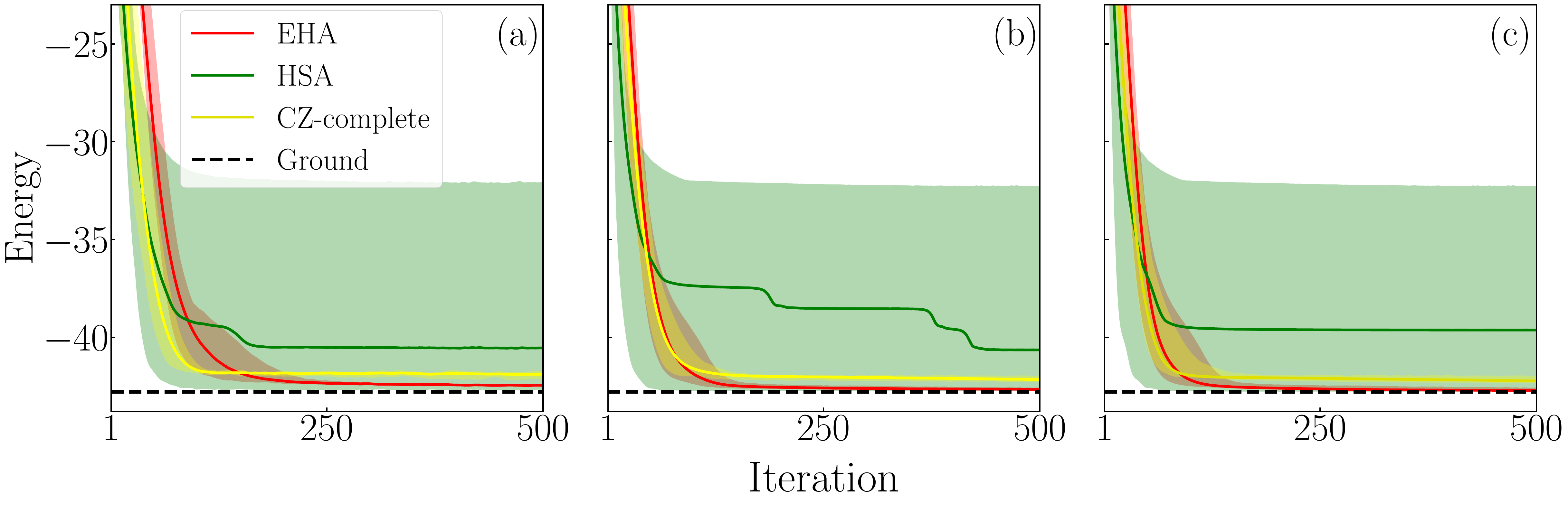}
	\caption{Performance of EHA, HSA and CZ-complete under different numbers of measurement shots for the TFIM in Eq.~(\ref{eq:TFIM}). (a) Number of shots is 100. (b) Number of shots is 1000. (c) Number of shots is 5000. }
	\label{fig:TFIM_shot}
\end{figure}

We consider the practical case where the measurement shots are limited when evaluating the expectation of the Hamiltonian with respect to the output states of the circuits.  We take the TFIM in Eq.~\eqref{eq:TFIM} as an illustration. Here, we only compare the performance of three typical ansatzes: CZ-complete, HSA and our EHA. The experimental settings are the same as those in Fig.~\ref{fig:TFIM_cost}, except for the number of measurement shots. For the ideal case in Fig.~\ref{fig:TFIM_cost}, the shots are essentially infinite. Here, we consider three different numbers of shots to evaluate the expectation of the Hamiltonian, namely, 100, 1000 and 5000. The experimental results are demonstrated  in Fig.~\ref{fig:TFIM_shot}. It is clear that our EHA outperforms HSA and CZ-complete.

\section{RESOURCES FOR CHEMICALLY INSPIRED ANSATZES}
\label{app:chem}

\begin{table*}[h]
	\begin{ruledtabular}
	\begin{tabular}{cccccccc}
    Ansatz     & Molecule                  & All CX gates & Non-local CX gates & T gates \\
	\hline
	UCCSD      & $\mathrm{H_3}^{+}$(1.1)            & 272        & 32                   & 0  \\
	UCCSD      & HF(1.1)                   & 2400       & 320                  & 0  \\
	GRSD       & $\mathrm{H_3}^{+}$(1.1)            & 64         & 45                   & 16 \\
	GRSD       & HF(1.1)                   & 370        & 272                  & 40 \\
    ADAPT-VQE  & $\mathrm{H_5}$(*)         & 306        & 246                  & 24 \\
	ADAPT-VQE  & BeH2(1.1)                 & 446        & 342                  & 24 \\
	ADAPT-VQE  & LiH(1.11)                 & 366        & 283                  & 32 \\
	ADAPT-VQE  & HF(1.1)                   & 152        & 109                  & 24 \\
	\end{tabular}
	\end{ruledtabular}
	\caption{\label{tab:resource}Resources for different chemically inspired ansatzes. }
\end{table*}
Here, for the $\mathrm{H_5}$ molecule,  (*) represents all bond lengths ranging from 0.5 to 1.5 with a step size of 0.1 (see Fig.~\ref{fig:H5}).

\section{THE IMPACT OF INITIAL REFERENCE STATES FOR HSA}
\label{app:HSA}

\begin{figure}[H]
	\centering
	\includegraphics[width=0.95\textwidth]{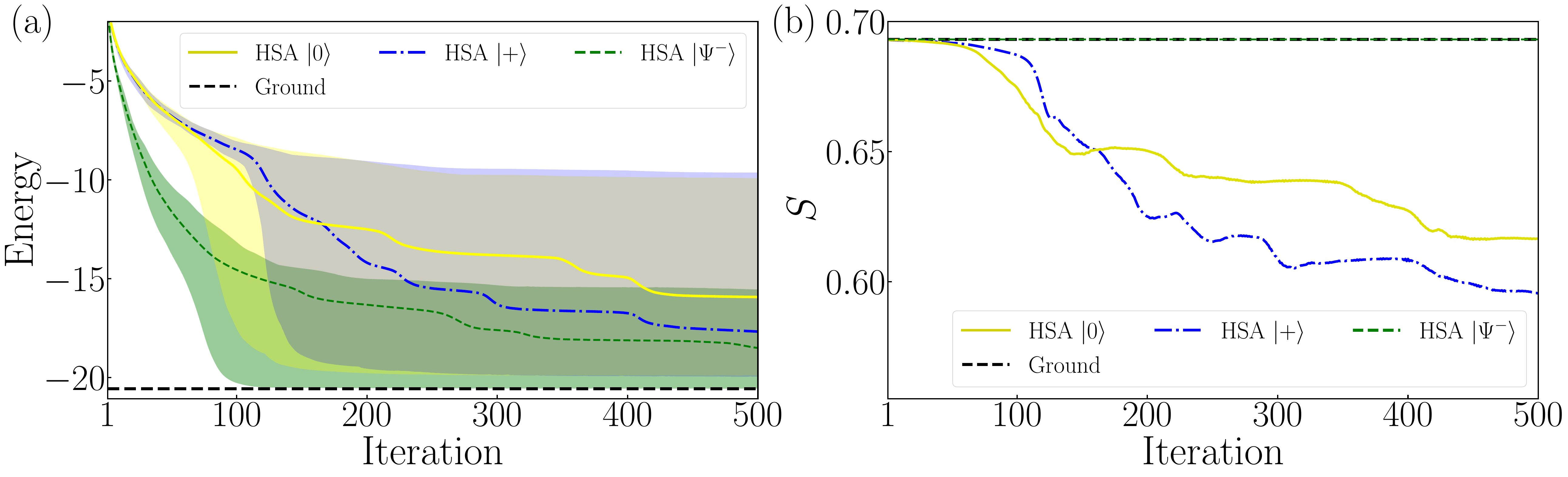}
	\caption{Performance of HSA under different initial reference states for HM in Eq.~\eqref{eq:XXZ}. (a) The black dashed line denotes the actual ground state energy,
and all the other lines indicate the respective average value of the energy over 10 realizations under different initial reference states. The shaded areas represent the smallest area that includes all the behaviors of the 10 realizations. (b) The black dashed line denotes the actual entanglement of the ground state, and all the other lines indicate the respective average value of the entanglement over 10 experiments under different reference states. }
	\label{fig:HSA_init}
\end{figure}

From Fig.~(\ref{fig:HSA_init})(a), we find that for the HM, under the reference states ${\otimes^{12}}\ket{+}$ and ${\otimes^{12}}\ket{0} $, HSA cannot converge to the ground state. This validates that the performance of the problem-specific HSA is largely dependent on the reference state.
This phenomenon can be explained by checking the corresponding  entanglement transitions in Fig.~(\ref{fig:HSA_init})(b). We find that once the initial state is changed to ${\otimes^{12}}\ket{+}$ or ${\otimes^{12}}\ket{0}$, HSA cannot maintain the same amount of entanglement as the ground state, and in turn resulting in poor performance.

\section{STEP SIZE SCHEDULE}
\label{app:step}
\begin{table}[htpb]\label{stepsize1}
	\begin{ruledtabular}
	\begin{tabular}{cccccccc}
		&Model & Step size  & Iteration steps\\
		\Xhline{0.5\arrayrulewidth}
		&$\mathrm{BeH}_2$(1.1) &0.01  &1000\\
		&                      &0.005 &2000  \\
		\Xhline{0.5\arrayrulewidth}
		&LiH(1.11)             &0.01  &1000  \\
		&                      &0.005 &1000  \\
		\Xhline{0.5\arrayrulewidth}
		&HF(1.1)               &0.01  &1000  \\
		\Xhline{1.5\arrayrulewidth}
		&HM(8)               &0.01  &1000 \\
		\Xhline{0.5\arrayrulewidth}
		&HM(12)               &0.005  &1000 \\
		&                     &0.001 &1000 \\
		&                     &0.0005 &2000\\
		\Xhline{0.5\arrayrulewidth}
		&HM(16)               &0.005  &3000 \\
		&                     &0.001 &2000 \\
		&                     &0.0005 &2000 \\
		\Xhline{0.5\arrayrulewidth}
		&TFIM1(8)               &0.01  &2000 \\
		\Xhline{0.5\arrayrulewidth}
		&TFIM1(12)              &0.01 &4000 \\
		\Xhline{0.5\arrayrulewidth}
		&TFIM1(16)             &0.01 &4000  \\
		\Xhline{0.5\arrayrulewidth}
		&TFIM2(8)               &0.05  &500 \\
		&               &0.02  &1000 \\
		\Xhline{0.5\arrayrulewidth}
		&TFIM2(12)              &0.02 &1500 \\
        &                        &0.01 &2000 \\
		\Xhline{0.5\arrayrulewidth}
		&TFIM2(16)             &0.02 &1000  \\
		&                       &0.01 &4000  \\
		\Xhline{0.5\arrayrulewidth}
		&BHM(8)               &0.01  &1000 \\
		\Xhline{0.5\arrayrulewidth}
		&BHM(16)               &0.01  &500 \\
		&                     &0.005  &500 \\
	\end{tabular}
	\end{ruledtabular}
	\caption{\label{tab:iter_schedule} The step size schedule for different models.}
\end{table}

\end{document}